\documentclass[11pt]{article}
\usepackage[a4paper, margin=1in]{geometry}

\usepackage{authblk}

\usepackage[T1]{fontenc}
\usepackage{graphicx,url}
\usepackage{color, xcolor}
\usepackage{graphicx}
\usepackage[labelformat=simple]{subcaption}
\usepackage{xspace}
\usepackage{multirow}
\usepackage{url,xurl}
\usepackage{comment}
\usepackage{colortbl}
\usepackage{amsfonts}
\usepackage[hidelinks]{hyperref}
\usepackage{amsmath}
\usepackage[capitalise]{cleveref}
\usepackage{longtable}
\usepackage{booktabs}  
\usepackage{array}  
\usepackage{float} 
\usepackage{ifthen}
\usepackage{enumerate}
\usepackage{enumitem}

\usepackage{footnote}
\makesavenoteenv{longtable}
\usepackage{threeparttable}
\usepackage{tablefootnote}
\usepackage[title]{appendix}

\newcommand\cmt[1]{\ifthenelse{\boolean{includecomments}}{{\color{blue} \sffamily [Comment: #1]}}{}}
\newcommand\leandro[1]{\ifthenelse{\boolean{includecomments}}{{\color{orange} \sffamily [Leandro: #1]}}{}}
\newcommand\pedro[1]{\ifthenelse{\boolean{includecomments}}{{\color{red} \sffamily [Pedro: #1]}}{}}
\newcommand\renan[1]{\ifthenelse{\boolean{includecomments}}{{\color{green} \sffamily [Renan: #1]}}{}}
\newcommand\gabriela[1]{\ifthenelse{\boolean{includecomments}}{{\color{purple} \sffamily [Gabriela: #1]}}{}}
\newcommand\reviewfix[1]{\ifthenelse{\boolean{includecomments}}{{\color{red} \sffamily [ReviewFix: #1]}}{}}

\title{Analyzing the Effect of an Extreme Weather Event on Telecommunications and Information Technology:\\ Insights from 30 Days of Flooding\thanks{This is a preprint of the paper accepted for publication at Passive and Active Measurement Conference (PAM) 2025, published in Lecture Notes in Computer Science (LNCS), Springer. The final authenticated version is available online at \url{https://doi.org/10.1007/978-3-031-85960-1_12}.}}

\author[1]{Leandro Márcio Bertholdo}
\author[2]{Renan Barreto Paredes}
\author[3]{Gabriela de Lima Marin}

\author[4]{Cesar A. H. Loureiro}
\author[3]{Milton Kaoru Kashiwakura}
\author[2]{Pedro de Botelho Marcos}

\affil[1]{Federal University of Rio Grande do Sul, Brazil\\
}
\affil[2]{Federal University of Rio Grande, Brazil\\
}
\affil[3]{Brazilian Network Information Center (NIC.br), Brazil\\
}
\affil[4]{Federal Institute of Rio Grande do Sul, Brazil\\
}

\date{} 

\newboolean{includecomments}
\setboolean{includecomments}{false} 

\begin{document}
	\maketitle

\begin{abstract}

In May 2024, weeks of severe rainfall in Rio Grande do Sul, Brazil caused widespread damage to infrastructure, impacting over 400 cities and 2.3 million people. 
This study presents the construction of comprehensive telecommunications datasets during this climatic event, encompassing Internet measurements, fiber cut reports, and Internet Exchange routing data.
By correlating network disruptions with hydrological and operational factors, the dataset offers insights into the resilience of fiber networks, data centers, and Internet traffic during critical events.
For each scenario, we investigate failures related to the Information and Communication Technology infrastructure and highlight the challenges faced when its resilience is critically tested.
Preliminary findings reveal trends in connectivity restoration, infrastructure vulnerabilities, and user behavior changes. These datasets and pre-analysis aim to support future research on disaster recovery strategies and the development of robust telecommunications systems.


\end{abstract}

\section{Introduction} \label{sec:introduction}

In May 2024, the state of Rio Grande do Sul, located in southern Brazil, experienced\ unprecedented severe weather events, resulting in damage to 84\% of its cities \cite{defesacivil_municipios}. This major climatic event caused various disruptions to the state's infrastructure, including harm to roads, bridges, electrical plants, data centers, communication systems, and houses, affecting the lives of millions of people. According to the state's Civil Defense Department, displaced individuals exceeded 300,000 and affected 2.3 million people \cite{defesacivil_balanco}. Economic activity in the area was reduced by 94\%, affecting commerce and industry \cite{fiergs2024industrias}.

In disaster scenarios such as this, communication systems play a critical role for several reasons. First, they facilitate coordination among rescue teams, enabling them to locate individuals needing assistance. Second, they allow individuals to communicate with family and friends, informing them of their safety and sharing crowd-sourced information, such as reports of blocked roads and damaged bridges. Finally, communication systems support the continuity of public services and economic activities, helping to maintain order and prevent looting in stores and supermarkets.

During this event, Rio Grande do Sul's communication systems and infrastructure were severely strained as some of its elements were partially or entirely disrupted by the consequences of the climate event. In this paper, we investigate how these events affected the resilience of the infrastructure and operations of the communications systems in Rio Grande do Sul. We analyze data from multiple sources and provide insight to improve ICT infrastructures.


We organize this paper as follows. Section~\ref{sec:events} outlines the climate events and their impact on ICT infrastructures. Sections~\ref{sec:aerial} through \ref{sec:rnp} examine disruptions to fiber networks, data centers, and long-distance circuits. Sections~\ref{sec:ix} and \ref{sec:simet} analyze the impact on the regional Internet Exchange (IXP) and the perceived impact on Internet quality for end users, respectively. Section~\ref{recovery} assesses recovery efforts, while Section~\ref{sec:related} compares this case with other weather-related events. Section~\ref{sec:lessons} discusses lessons learned, and Section~\ref{challenges} highlights open challenges and future research directions.

\section{Background and Sequence of Events}
\label{sec:events}

The unprecedented severe weather events of May 2024 caused catastrophic damage across 418 municipalities in Rio Grande do Sul, significantly impacting the state capital, Porto Alegre, and its metropolitan area. The civil defense classified the event as a Level III on a scale of four \cite{defesacivil_municipios}. The most recent flooding event of similar magnitude occurred 83 years ago, in 1941, leading to the city constructing a flood protection system.

\begin{figure}
	\centering
	\includegraphics[width=1\textwidth]{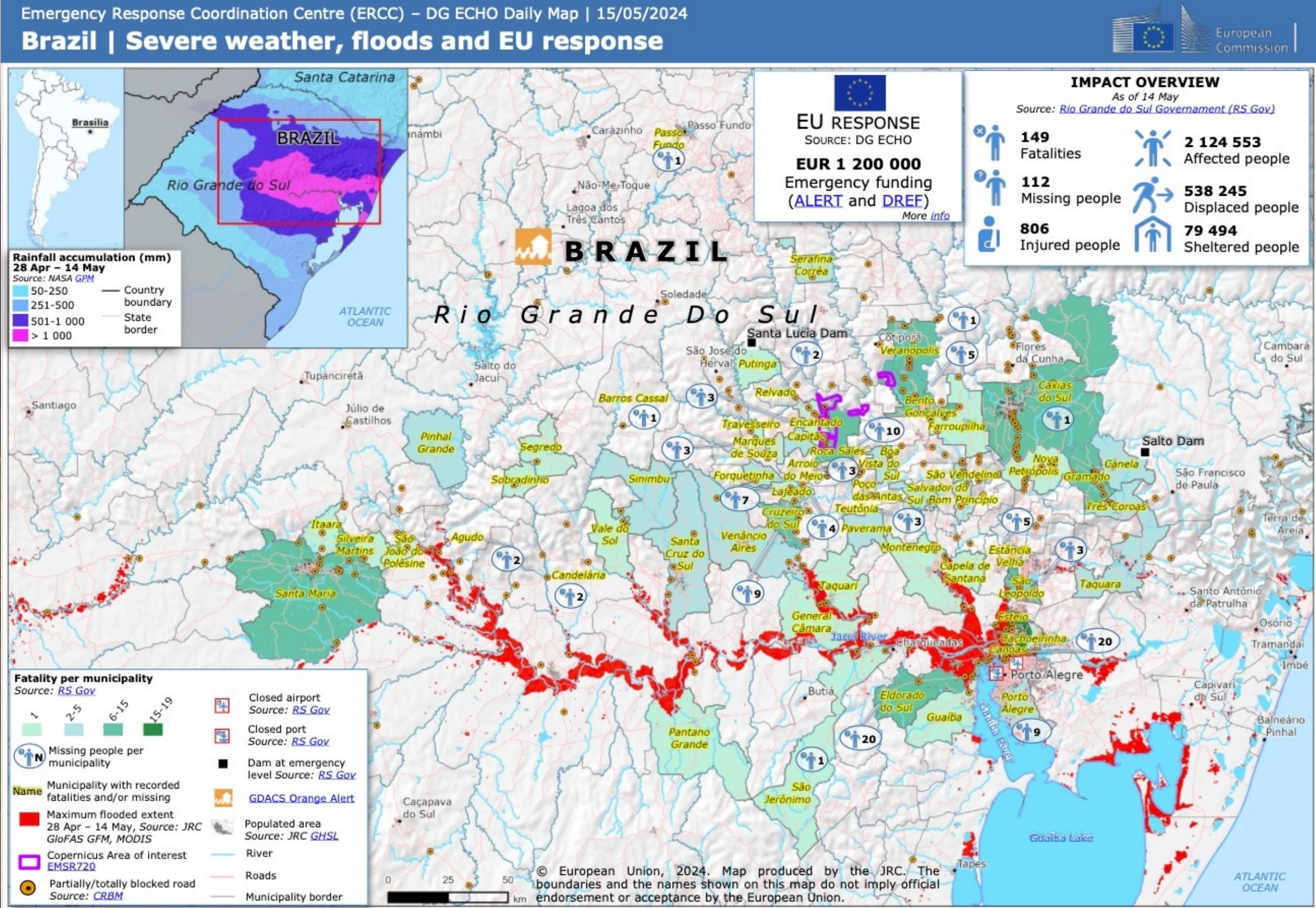}
	\caption{Map of the areas affected by the 2024 floods in the state of Rio Grande do Sul. Source: ERCC GPM, Rio Grande do Sul Government, Copernicus EMSR. \protect\reviewfix{A1} }
	\label{fig:affected-areas-RS}
\end{figure}

The disaster began with heavy rainfall on April 27 and intensified on April 29. The volume of rainfall surpassed 1,000 mm in the region (1 meter of water height per square meter of land) \cite{metsul1000mm}, significantly impacting the valleys of the Taquari, Caí, Jacuí, Sinos, and Gravataí rivers. All these rivers are part of the so-called Gua\'{i}ba Basin, flowing into the Gua\'{i}ba River, which runs through the capital, Porto Alegre. The Gua\'{i}ba River flows first into the Lagoa dos Patos and then to the ocean. The Lagoa dos Patos is the largest lagoon in South America, with a length of 265 kilometers (map on \autoref{fig:affected-areas-RS}).

In smaller cities, accumulated rainfall was the primary cause of flooding, while in the metropolitan area and the capital, natural and human factors contributed to it. In addition to the high rainfall, strong winds prevented the water from Lagoa dos Patos from flowing into the ocean, creating a natural blockage and leading to excessive water accumulation in the Gua\'{i}ba River, which caused prolonged flooding in the metropolitan area \cite{g1_baciaguaiba}. 
Furthermore, decades of neglected maintenance and updates to the Porto Alegre flood protection system increased the impacts of the disaster \cite{dmae_falha}. The flood protection system comprises 68 km of dikes, a protective wall, 23 pumping stations, and 14 floodgates \cite{ufrgs2024cedep}. \reviewfix{A5}

\begin{figure}
	\centering
	\includegraphics[width=1\textwidth]{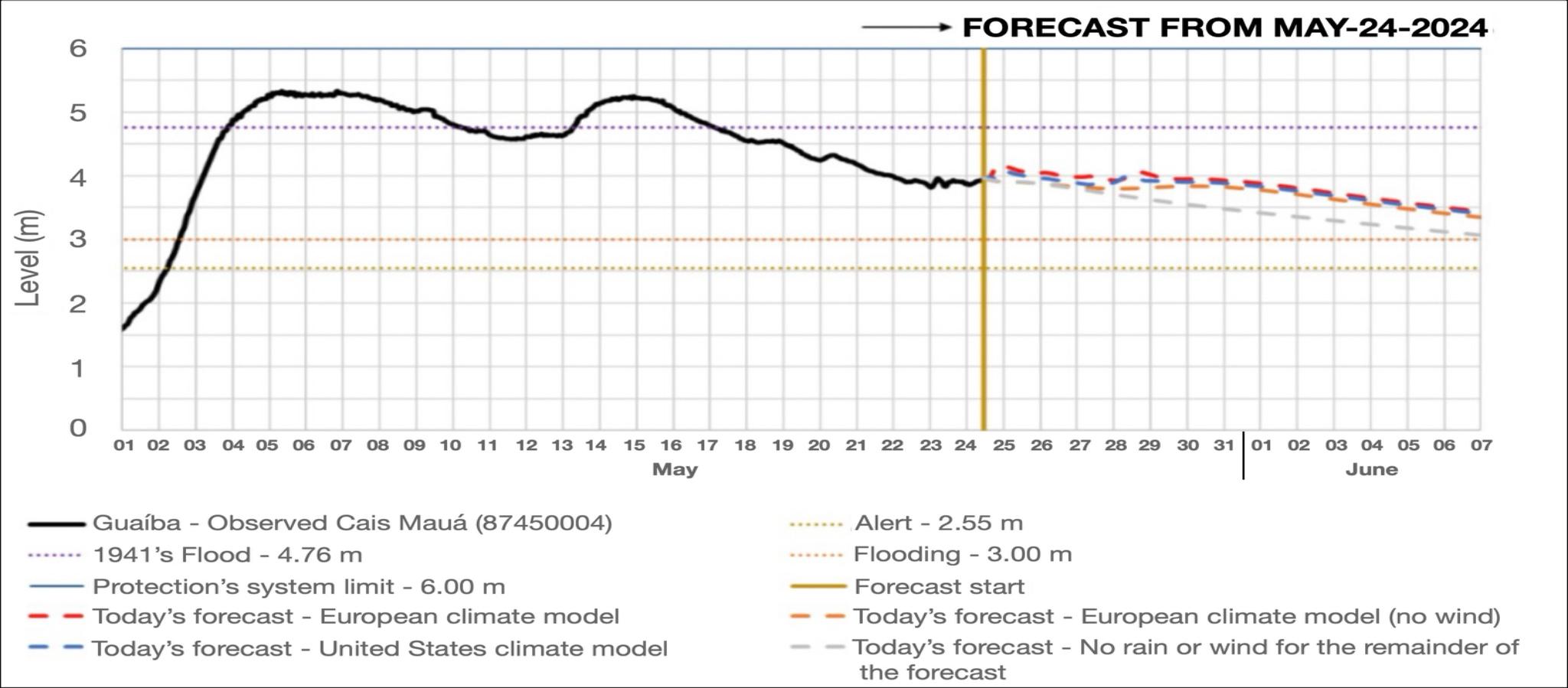}
	\caption{Gua\'{i}ba water level measurement and forecast~\cite{previsaoIPH}.\protect\reviewfix{A1} }
	\label{fig:iphufrgs}
\end{figure}

The bad weather persisted for a month. During this time, forecast models from the Institute of Hydraulic Research at the Federal University of Rio Grande do Sul (IPH/UFRGS) accurately predicted the situation several days in advance. Using US (dashed blue) and EU (red) weather models, they projected a return to normal water levels by mid-June. \autoref{fig:iphufrgs} illustrates the water level behavior during this period. The Gua\'{i}ba River receded below the alert level on June 7th, but issues in the city's drainage system prolonged water retention in some areas. 


This disaster underscores the severity of the climatic event and the extensive disruptions it caused to infrastructure and the lives of people in Rio Grande do Sul. The timeline of the disaster is presented in \autoref{tab:flooding_events}.

\begin{center}
	\footnotesize
	\begin{longtable}{>{\raggedright}p{2.2cm} >{\raggedright\arraybackslash}p{0.75\linewidth}}
		\caption{Significant events related to the flooding in Porto Alegre. ICT infrastructure marked.}
		\label{tab:flooding_events}
		\\
		\toprule
		\textbf{Date} & \textbf{Event} \\
		\midrule
		\endfirsthead
		
		\multicolumn{2}{c}{\small\itshape (continued) Significant events related to the flooding in Porto Alegre} \\
		\toprule
		\textbf{Date} & \textbf{Event} \\
		\midrule
		\endhead
		
		\midrule
		\multicolumn{2}{r}{\small\itshape Continued on next page} \\
		\endfoot
		
		\bottomrule
		\endlastfoot
		
		\rowcolor{lightgray!50}April 27 & The first storm impacts the Vale do Rio Pardo region~\cite{cronologia2024g1}. Over 1{,}200 lightning strikes were recorded in Porto Alegre, making fiber cable maintenance unfeasible due to the risk of electrical discharge. \\
		April 28 & The National Institute of Meteorology (Inmet) issues an orange alert for storms across the state's southern half.  \\
		\rowcolor{lightgray!50}April 29 & The energy grid is affected, and the shortage is a consequence of the heavy rain, wind, and hail recorded over the weekend. Several communities are isolated. Inmet has issued a red alert, indicating the onset of a major disaster.\\
		April 30 & The first deaths are reported, roads blocked, and bridges washed away. Gua\'{i}ba River floods riverside areas and a state of emergency is declared. \\
		May 1 & Heavy rains cause landslides and widespread flooding, prompting evacuations in high-risk areas. \\
		\rowcolor{lightgray!50}
		May 2 & The last RNP fiber optics backbone redundancy failed, isolating the state's academic network. The Gua\'{i}ba River reached a flood level of 3.60 meters for the first time. \\
		May 3 & Severe impacts in Porto Alegre include major transportation disruptions and infrastructure damage. Gua\'{i}ba reaches 4.77 meters, breaking the 1941 record. Salgado Filho Airport closes, and communities in seven cities are evacuated.\\
		\rowcolor{lightgray!50}
		May 4 &The incumbent data center (ELEA-POA2) is shutdown~\cite{elea2offline2024}. \reviewfix{C4}\\
		May 5 &The Gua\'{i}ba River continues to rise, reaching a new historical peak of 5.35 meters. Main roads to Porto Alegre interrupted. \\
		\rowcolor{lightgray!50}
		May 6 & Mobile operators enabled free roaming on 3G, 4G, and 5G networks~\cite{roaming_celular}. The PROCERGS data center was shut down~\cite{procergs-offline2024cpovo}, and the Scala data center ceased operations after three days on generators~\cite{scala2024pressrelease}. The government granted ISPs access to emergency roads, restricted to rescue operations, for cable repairs~\cite{internetsul_dnit}. \reviewfix{A2}\\
		\rowcolor{lightgray!50}
		May 7 & Long-distance circuit failures registered around 20\% at PoP-RS/RNP, isolating several universities, hospitals, and research centers.\\
		\rowcolor{lightgray!50}
		May 9 & PROCERGS, PoP-RS/RNP, and CPFL-T share fiber and infrastructure in Porto Alegre to relocate the state government's crisis office. \\
		\rowcolor{lightgray!50}
		May 20 & Riots involving burning public transport buses caused new fiber cuts~\cite{GZH2024onibus}.\\
		June 1 & The Gua\'{i}ba River level fell below the flooding level~\cite{abaixocota1jun}.\\
		June 3 & Gua\'{i}ba River level rises again, surpassing the flood level~\cite{guaibavoltasubir}.\\
		June 7 & The Gua\'{i}ba River returns to its banks for the first time in a month~\cite{guaibaretornaleito}.\\
		
	\end{longtable}
\end{center}

\section{Optical Fiber Networks}
\label{sec:aerial}

To investigate the impacts on optical fiber networks, we analyzed data from two backbone networks: an academic network (Metropoa) and an Internet Service Provider (ISP) backbone. The Metropoa network \cite{site_metropoa} is a 190 km optical fiber infrastructure in the Porto Alegre metropolitan area, connecting approximately 180,000 individuals across educational institutions, research centers, and university hospitals. Approximately 85\% of its fiber paths are aerial, with the remaining 15\% underground. The ISP operates a 500 km backbone network and an additional 2,000 km of last-mile infrastructure, serving 60,000 households, 20,000 of which were affected during the event. \reviewfix{D1,D2}


In Brazil, most fiber is aerial, installed at a height of 4 meters on utility poles, making it highly susceptible to climate events. During the incident, rescue teams intentionally cut many cables to allow the passage of high trucks and boats carrying rescued individuals, preventing accidents. Additionally, there was an increase in cable cuts for copper theft, often involving loose cables left hanging on power poles. Field technicians distinguish stolen cables, identified by straight cuts, from naturally broken cables, which typically show unevenly torn internal fibers. This information is documented in the Metrofiber datasets~\cite{github2024furg}, which offer opportunities for further analysis and insights.\reviewfix{A3}

Here, we correlate a series of events affecting the Metropoa network over time and map the impact of climate-related events on this optical network in recent years. Both direct and indirect climate-related events are considered. Direct events include tree falls on cables during storms or cyclones, while indirect events involve cable breakages caused by fires triggered by damage to the electrical network during rainstorms. \reviewfix{Rephrased for clarity and conciseness.}

\begin{figure}[htb!]
	\centering
	\includegraphics[width=1\textwidth]{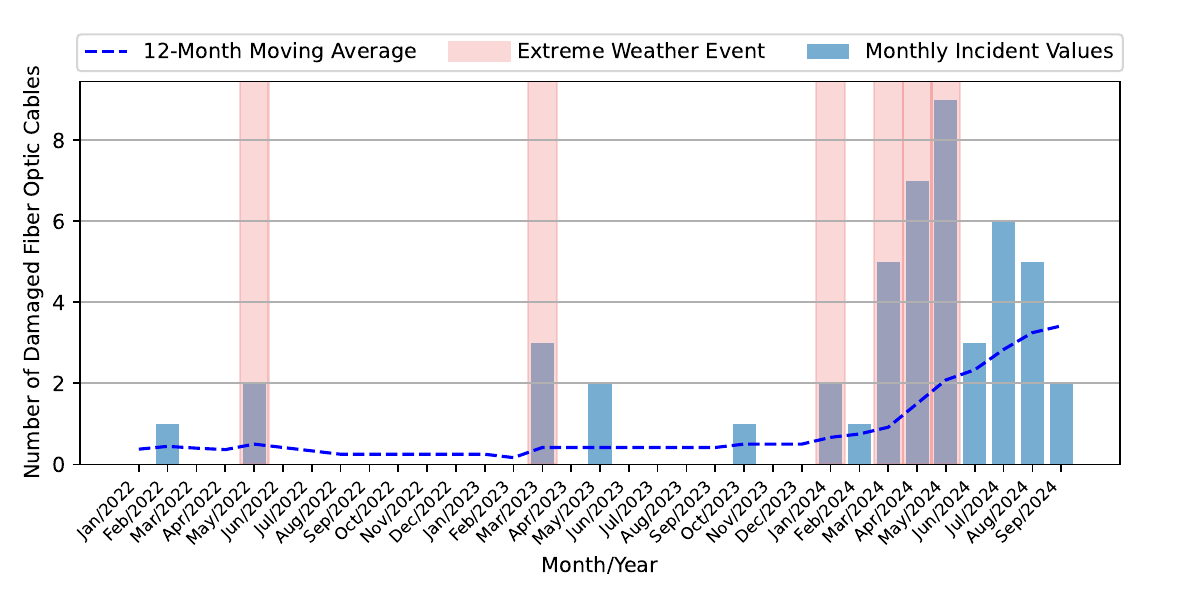}
	\caption{Optical cable cuts and weather-related events in a fiber network backbone.}
	\label{fig:metropoa-incidents}
\end{figure}

In \autoref{fig:metropoa-incidents}, we illustrate the impact of climate-related events on the number of cable damages in the aerial portion of the academic network between 2022 and 2024. These events include Cyclone Yakecan in May 2022 \cite{temporal_mai2022globo}, thunderstorms in March 2023 \cite{temporal_mar2023gzh}, January 2024 \cite{temporal_jan2024metsul}, and March 2024 \cite{temporal_mar2024cnn}, as well as a significant flooding event in May 2024 \cite{brazil_heavy_rains2024bbc} and the subsequent rebuilding process. Similar results for the ISP backbone are presented in Appendix.


The bad weather also increased the mean time to repair aerial cable cuts, as electrical storms made climbing poles to fix cables unsafe. This situation typically delayed repairs by one or two days. However, aerial networks made it possible to create new emergency paths or reconnect customers and servers by moving their ICT infrastructure to other locations, as the underground fiber was hard to access in these cases.
On the other hand, no damage was recorded in the underground fiber network analyzed (two events in the last 12 years on 28 km of fiber). The reliability of the underground network proved advantageous, providing a secure path that enabled the network owner to offer fiber routes to institutions whose networks or data centers were damaged. By ``reliable'',  we refer to a cable with all fiber routes intact, \textit{i.e.}, all fibers fused in each splice box. Typically, providers only fuse active fibers, leaving unused fibers unfused in splice boxes to reduce repair costs.\reviewfix{A4}

Following the 2022 cyclone, the municipality of Porto Alegre enacted a law requiring all energy and telecommunication cables to be installed underground within 15 years~\cite{lei137070poa}. The new law was approved in May 2023, a year before the major flooding event. The Internet Providers Association estimates that underground installations are 50  times more expensive than aerial ones, with added challenges such as higher costs for fault detection and repairs \cite{internetsul2023dutos}. Nationally, only 1\% of energy and fiber cables are underground, compared to 10\% in São Paulo and 9\% in Porto Alegre, where underground installations are estimated to be 15 times more expensive than aerial alternatives \cite{shared2022simao}.\reviewfix{A4}


\textbf{Takeaway}: Underground networks are more resilient to climate events than aerial ones. A resilient city should prioritize underground optical paths, ensuring all fibers, including unused ones, are pre-fused, and establish plans for alternative interconnections during emergencies.\reviewfix{A4}

\section{Data Centers in the Region}
\label{sec:datacenters}


In recent years, the municipality of Porto Alegre has incentivized businesses to establish operations in the 4th District, near the  Gua\'{i}ba River, by offering tax reductions \cite{4thdistrito}. This district, known for its proximity to the city center and airport, has become a hub for data centers, including ELEA-POA2, Scala Datacenter, and BRDigital Datacenter. Located along the banks of the Guaíba and Jacuí Rivers, these facilities were severely impacted by the flooding.

Starting on May 3, severe flooding in Porto Alegre caused the shutdown of several small and large data centers. A total of 21 out of the city’s 35 known facilities went offline. Notable affected centers included the Court of Justice of Rio Grande do Sul (TJRS)~\cite{tjrs-offline2024conjur}, the Federal Court of Justice for the 4th Region (TRF4)~\cite{trf4-offline2024convergenciadigital}, the Brazilian Micro and Small Business Support Service (SEBRAE-RS)\cite{sebrae-offline2024}, ELEA-POA2, the 
incumbent\footnotetext[1]{\label{incumbent}The company managing the former government-owned telecom infrastructure under a concession agreement.}\footnotemark[1] 
data center~\cite{elea2offline2024}, and Scala Datacenter, a Tier-3 facility hosting an interconnection for the Rio Grande do Sul IXP~\cite{scala2024pressrelease}. \reviewfix{C4}


The most significant incident occurred on May 6, 2024, involving the PROCERGS data center, a public facility responsible for data processing for the Rio Grande do Sul state government, located on reclaimed river land. Following a government decision to shut down a pumping station to mitigate the population's electric shock risk, the data center suffered extensive damage to its electrical systems and backup generators. Authorities shut down the facility to prevent further damage and potential data loss, leading to widespread outages that disrupted critical government services---such as online identity verification, tax processing, and social security payments. The main systems remained offline for over eleven days~\cite{procergs-offline2024baguete}.
The incident underscored the vulnerability of critical infrastructure to extreme weather events. \reviewfix{C4}

In contrast, there were notable examples of resilience and adaptability. The BRDigital data center, located 600 meters from the Gua\'{i}ba River, continued to operate successfully by transporting fuel via boats into the building for its generators (details in \cite{brdigital2024teletime}). The Bank of the State of Rio Grande do Sul (Banrisul) switched their operations to their second data center in the city, and the incumbent shifted 80\% of their operations from  ELEA-POA2 to ELEA-POA1, the last one located on higher ground. Data centers situated on high floors in flooded areas (\textit{e.g.,} the 25th floor), primarily used for radio and cellular network operations, faced unique challenges in maintaining fuel supplies. They relied on manual transport, boats, or external cranes to deliver fuel. After a week, these high-floor data centers were turned off.

\begin{figure}[ht!]
	\centering
		\includegraphics[width=.75\textwidth]{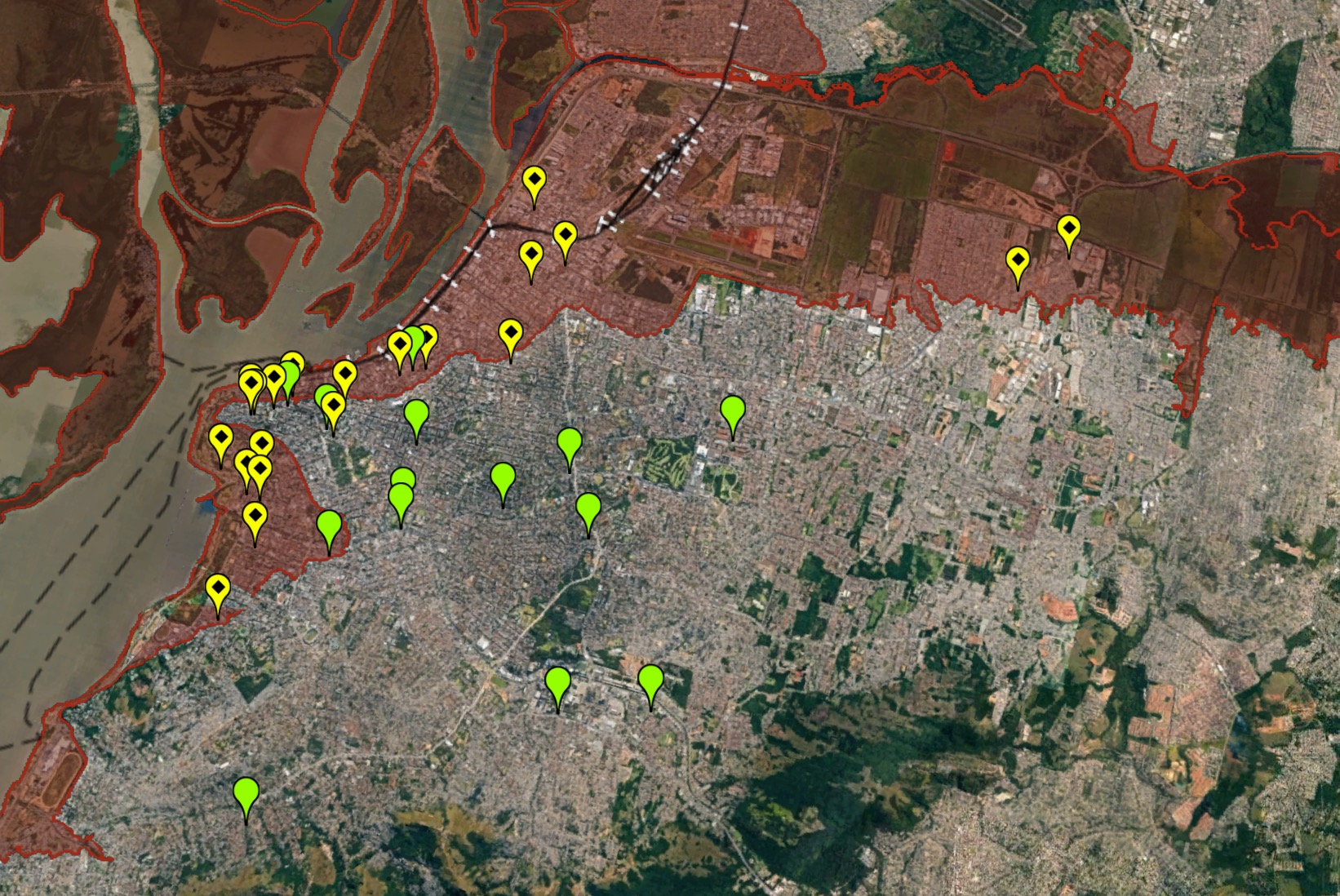}
	\caption{Map showing data centers in Porto Alegre. Yellow pins indicate data centers shut down due to floods. Green pins represent those that remained active. The red area shows the water level on May 6th (5.35m).}
	\label{fig:datacenter-flooding}
\end{figure}

Some other data centers near the flooded area were unaffected, such as the Federal University of Rio Grande do Sul and the municipality data processing center (PROCEMPA). In \autoref{fig:datacenter-flooding}, we show a map of all known data centers in the city that were either shut down or continued operating during the flooding.

In the days following, various levels of government accelerated their migration to cloud solutions \cite{cloud_procergs,cloud_tjrs}. Meanwhile, smaller companies sought to rescue their servers and data to temporarily host them with any operational collocation provider in the region, aiming to resume at least part of their operations. 

These responses highlight the importance of robust risk analysis in data center standards. The Brazilian NBR ISO/IEC 22237 series standards mention business risk analysis but do not specifically address flooding. The second part of this standard specifies minimum distances from specific locations such as airports, rivers, and gas stations. It is worth to mention some data centers are currently situated less than 100 meters from the Gua\'{i}ba River.

The ISO 27000 information security standard includes a specific table for environmental risks, including floods, where assets are located. This table classifies events as rare, frequent, and common, and these analyses are typically based on occurrences from the past 5 to 10 years to develop mitigation plans, not covering the last flooding. This recent event will likely impact new risk analyses, discouraging the construction of new data centers in those flooded areas.

\textbf{Takeaway}: During a 30-day flood, keeping electrical generators operational and fueled is challenging. Ensure backup data centers are 30 km away, preferably at higher elevations. This event should influence future risk analysis in this area.

\section{The National Research and Education Network (RNP) and Long-distance (WAN) Circuits}
\label{sec:rnp}

The Point-of-Presence of the National Research and Education Network in Rio Grande do Sul (PoP-RS) provides high-bandwidth Internet access for the academic and research community, including several major university hospitals. PoP-RS/RNP is hosted in the Federal University of Rio Grande do Sul Data Center (CPD-UFRGS), which remained unaffected during this critical period. In this section, we analyze how the long-distance connections from PoP-RS behaved during this time.

\begin{figure}[ht!]
	\centering
	\includegraphics[width=1\linewidth]{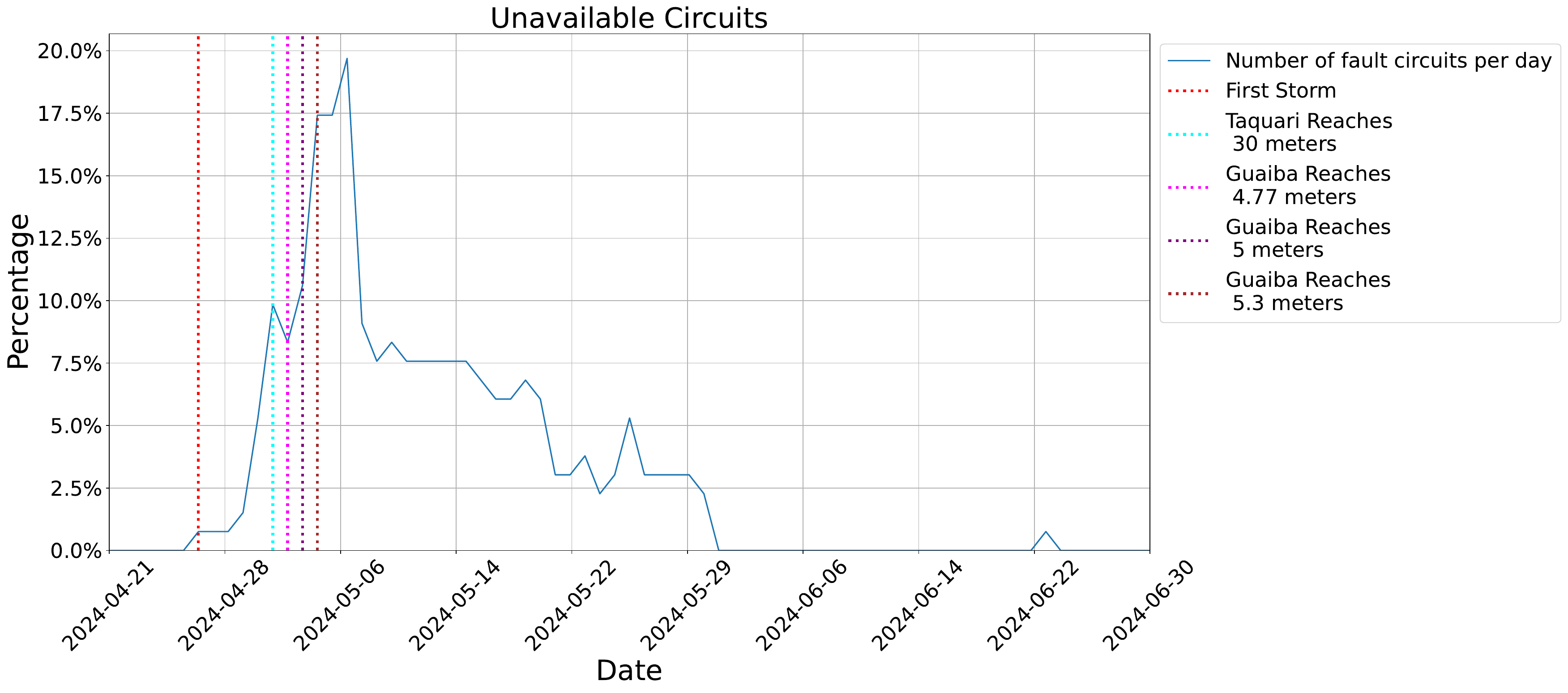}
	\caption{Percentage of long-distance (WAN) circuits down for more than 24h. at PoP-RS/RNP.}
	\label{fig:fault-circuits}
\end{figure}

Before the event, from April 20 to 26, no customer circuits were unavailable for more than 24 hours. There were isolated issues from April 27 to 30, but still without a direct relation to the flood. However, starting on May 1, the number of problems increased significantly, reaching 19.7\% unavailable circuits (see \autoref{fig:fault-circuits}). After May 31, the number of circuits unavailable for more than 24 hours had returned to normal. Upon deeper analysis, we found that these were institutions whose internal structures were affected by the flood. From May 1 to 31, 237 Internet circuit failures occurred in long-distance connections.

A second fact is related to long-distance circuits connecting the PoP-RS to the National RNP backbone and, consequently, to the Internet. The PoP-RS has four circuits of 100Gbps to the RNP backbone: two paths using OPGW (Optical Ground Wire) fiber via Eletrosul power towers connecting to two other state capitals, PR (Paraná) and SC (Santa Catarina); one circuit via the BRDigital telecom operator, also to the PR capital; and one circuit via the Latin American Cooperation Network (RedClara) that connects PoP-RS to SP (São Paulo) but with a high RTT (Round-Trip Time) of around 100ms, due to the cable route passing through Uruguay, Argentina and Chile before SP. 

During the climatic events on May 2, all backbone circuits were unavailable for 11 hours due to landslides in the Serra Gaúcha region. Partial restoration occurred after maintenance on one of the power towers, which remained at risk of collapse for 3 days. Despite having four distinct circuits, it is crucial to consider the complete path of the circuits. In the observed situation, all cables were close to BR-116, which crosses the Serra Gaúcha (highlands in the state heavily affected at the event's beginning). Full operation was normalized only after the BrDigital operator rerouted its circuit along the coastal route via BR-101.

Maintaining the RNP connection to the countryside and national and international Internet access supported various activities, from assisting the civil defense in predicting flooding effects in the city by connecting research groups from the university, such as the hydrology department at UFRGS responsible for water level projections, to allowing those sheltering inside university facilities around the state to communicate with their families through free internet access.

\textbf{Takeaway}: The academic infrastructure for Internet connections, as well as for other ISPs, has only three routes out of the state: the main roads BR-101 and BR-116, and the high-tension (OPGW) network. Thus, to increase resilience, telecommunications companies should create new paths.

\section{The Rio Grande do Sul Internet Exchange Perspective}
\label{sec:ix}

\begin{figure}[ht!]
	\centering
	\begin{subfigure}[b]{1\linewidth}
		\centering
		\includegraphics[width=.75\linewidth]{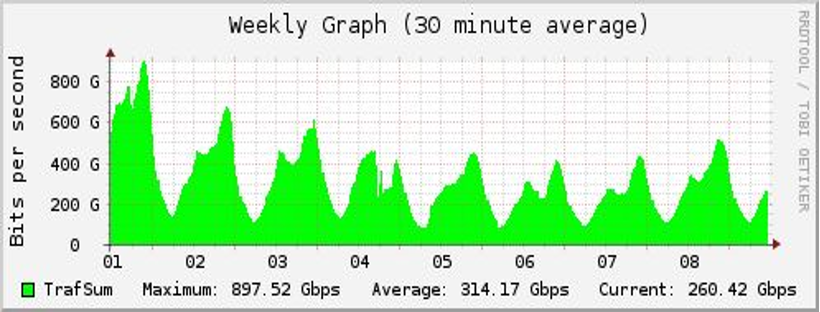}
		\caption{IX.RS experienced a traffic loss from May 1 to May 7, 2024.}

		\label{fig:rsix-1-9may}
	\end{subfigure}
	\hfill
	\begin{subfigure}[b]{.75\linewidth}
		\centering
		\includegraphics[width=\linewidth]{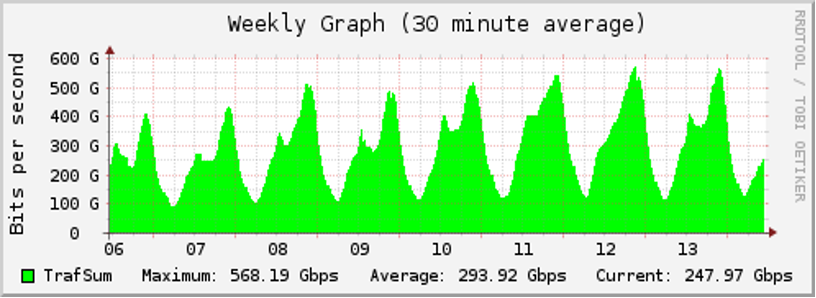}
		\caption{Traffic recovery of IX.RS after its lowest point on May 6.}
		\label{fig:rsix-06-13may}
	\end{subfigure}
	
	
	\caption{The chart illustrates the IXP's traffic loss from May 1-6, followed by a recovery beginning on May 7. Traffic peaked at approximately 400 Gbit/s on May 6-7 and stabilized around 600 Gbit/s after May 13. As of November 2024, the IXP has not fully recovered to its typical level of around 1 Tbit/s, and the exact reasons for the traffic not fully recovering are still under investigation. X-axis represents days in May 2024. \protect\reviewfix{C4}}
    \label{fig:rsix-comparison}
\end{figure}

During these events, many ISPs and data centers had their infrastructures affected. They lost connectivity to their upstream and the Rio Grande do Sul Internet eXchange Point~(IX.RS), leading to a 50\% decrease in traffic volumes~\cite{ixbr2024chuva}. \autoref{fig:rsix-comparison} compares the traffic volumes during the first two weeks of the events. According to operators from IX.RS, the increasing traffic volume in the second-week results from (sixteen) ISPs recovering their connectivity to the IXP.

In this section, we analyze routing data from IX-RS route servers (April 16 - May 24) to investigate the severe events' impact on connectivity and routes at the Porto Alegre IXP. A route server in an IXP is a critical component that simplifies and optimizes the exchange of routing information among participants~\cite{richter2014peering}. These data are part of the SARA project, a Portuguese acronym for Routing Analysis System for Autonomous Systems~\cite{sara_project}.\reviewfix{A6}

\begin{figure}[ht!]
	\centering
		\centering
		\includegraphics[width=1\textwidth]{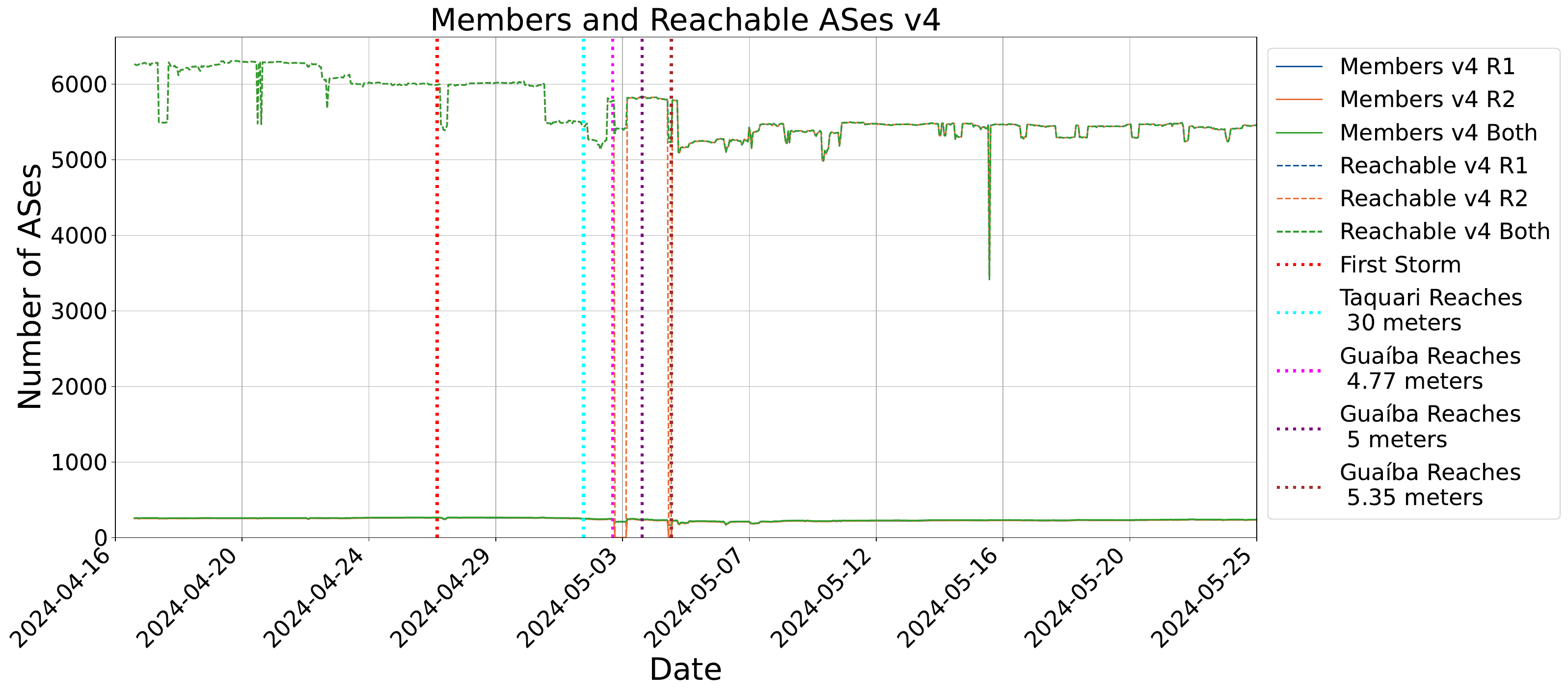}
	
	\caption{RSIX customer reachability.}
	\label{fig:main_IX}
\end{figure}

We first analyze the connectivity impacts of connected members, reachable ASes, announced prefixes, and available routes on IX-RS. We show these results in \autoref{fig:main_IX}, where we have the unfiltered results of the routing data from IX-RS. On April 27, considered the first storm of this event, the number of reachable ASes dropped around 10.1\% to 5389 ASes compared to April 26. Although it did recover, as the storms progressed and the water level kept rising, we saw a similar drop on April 30.


\begin{figure}[ht!]
	\centering
		\centering
		\includegraphics[width=1\textwidth]{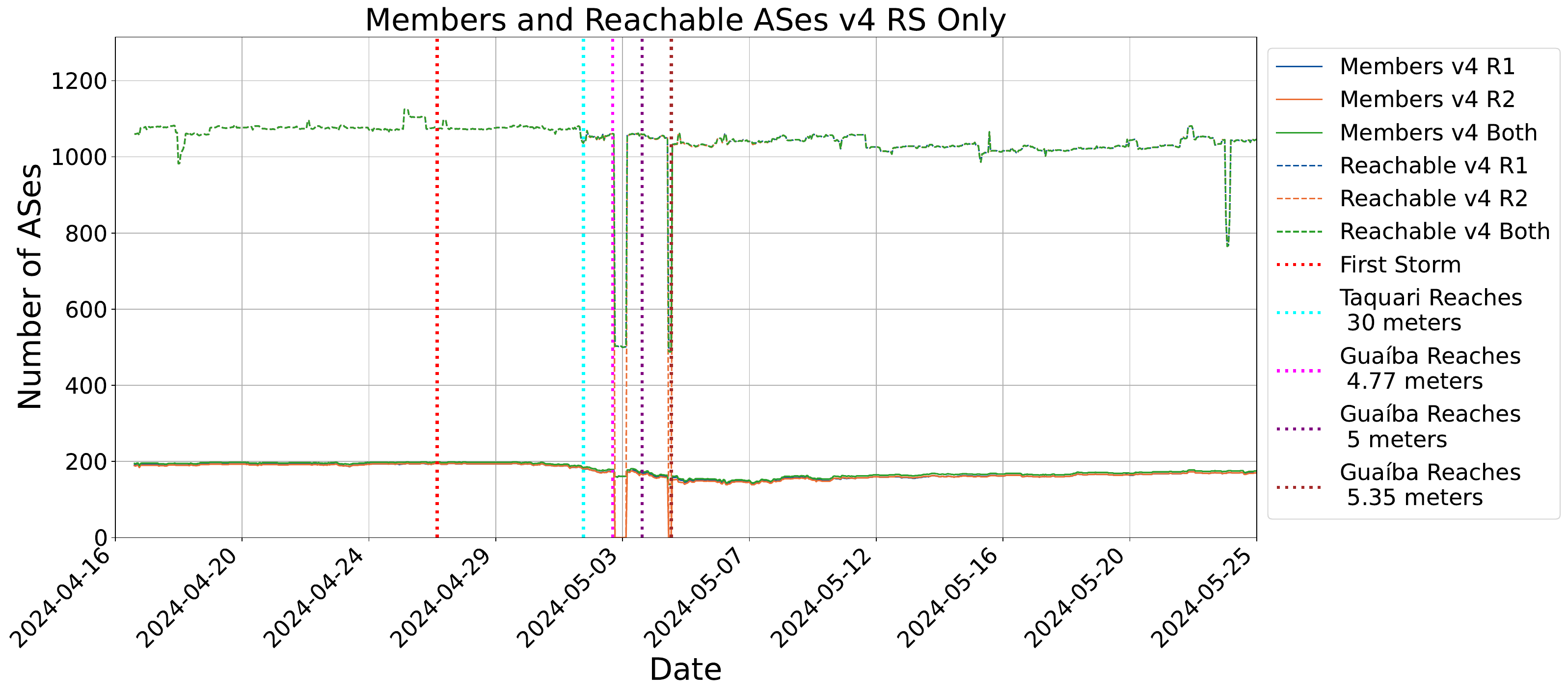}
	
	\caption{RSIX customer reachability filtered by geolocation.}
	\label{fig:RSonly_IX}
\end{figure}

During the flooding of Porto Alegre and the Taquari River, surpassing 30 meters, the number of reachable ASes and members significantly declined. On May 2, compared to April 26, reachable ASes dropped to 5,135 (a 14.4\% decrease), and members fell to 245 (an 11.2\% decrease). By May 5 at 05:51 AM UTC, these numbers declined to 5,093 reachable ASes (15.1\%) and 185 members (33\%). Notably, by May 24, the number of reachable ASes had not yet returned to regular levels. The drop in reachable ASes observed on May 15 is still under investigation. However, this drop did not involve ASes located in Rio Grande do Sul, as filtering the data for prefixes, members, or ASes in the state reveals no corresponding decline (\autoref{fig:RSonly_IX}). \reviewfix{C4}

By filtering to prefixes, members, or reachable ASes geolocated in the state of Rio Grande do Sul, we consider only instances where a member is located in the state or can reach a prefix, or AS, located in the state. With these criteria, we evaluate the connections provided by members in the state and the connections towards ASes in the state.
When filtering the data, we observe that the first storm did not significantly impact the number of members or reachable ASes that meet the mentioned criteria.
A decline can be seen, culminating in a sharp drop of members and reachable ASes on May 3rd, down 17\% and 53\%, respectively, together with Route Server 2 failing. The same happens again on May 4th, although in a shorter period, where we see a decline of 27.8\% of members and 54.6\% of reachable ASes.
By May 24th, the number of ASes had not recovered, and they even suffered another drop on May 23rd due to another rainfall.

\begin{figure}[h!]
	\centering
	\includegraphics[width=1\textwidth]{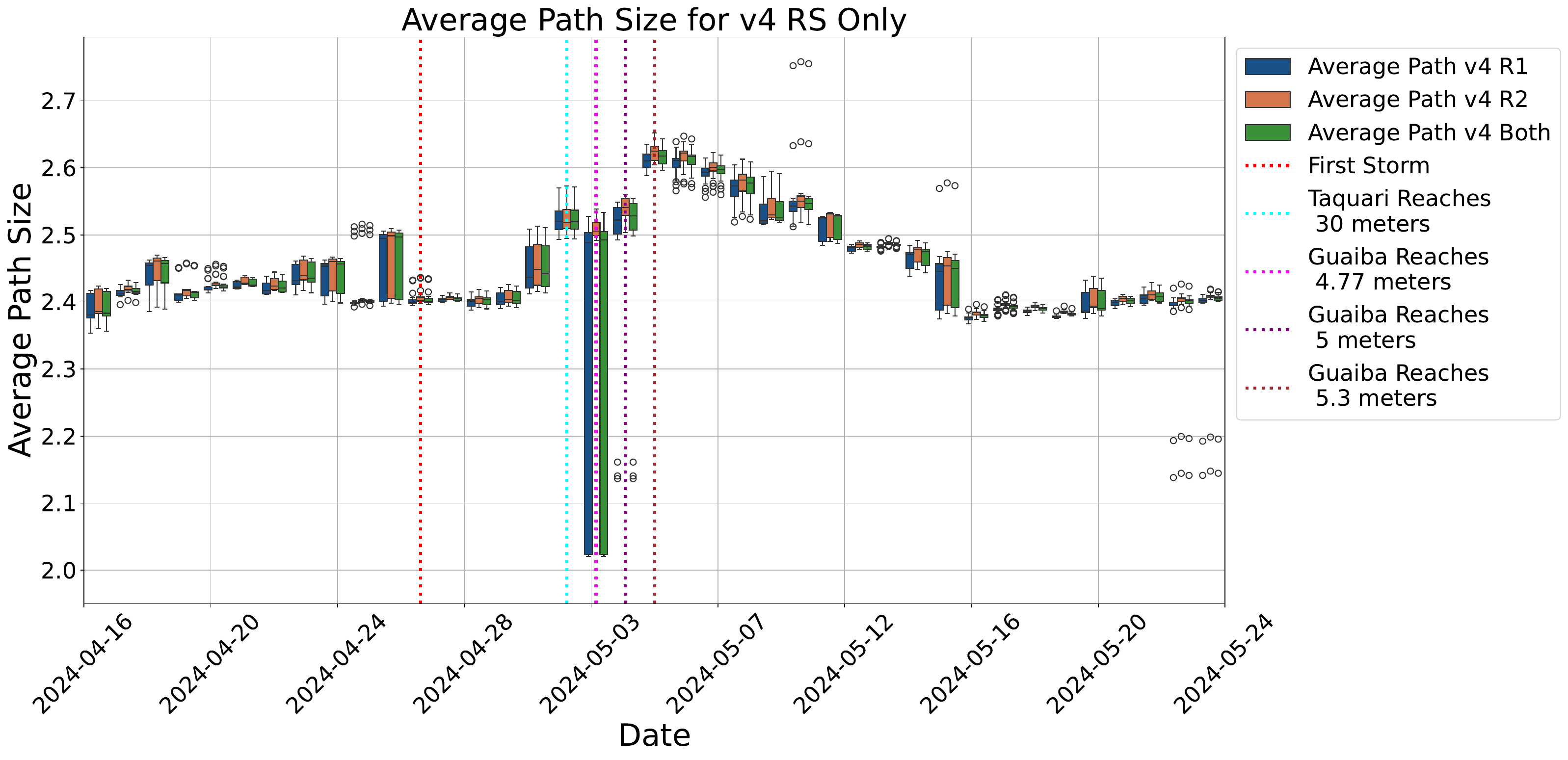}
	\caption{Average path size on Rio Grande do Sul Internet Exchange during the  critical period.}
	\label{fig:mediacaminhosrs}
\end{figure}

We also analyze the average path size of the routes announced at IX-RS of ASes from Rio Grande do Sul during the flooding period. We show these results in \autoref{fig:mediacaminhosrs}. We can notice that as the flooding propagates through the state, the path to each remaining network increases its average length (IPv4). 

Also, the first sharp decline in ASes, when the Gua\'{i}ba River reached 4.77 meters, reflected on the average path, especially in the 25th percentile, suggesting that the aforementioned decline affected routes with longer paths. 
The average path length started decreasing on the following days; by May 16, it had returned to similar levels before the floods began.

Despite the observable reduction in the number of connections and an increase in average path length, the IX-RS effectively maintained connectivity between ASes within Rio Grande do Sul and external ASes. This can be seen on May 7, when the number of members decreased by 25\% while the number of reachable ASes only dropped 3\%, taking into account both ASes (or prefixes) within the state and ASes connected through paths to those located in the state. 
It is important to note that the redundancy of route servers played a critical role in mitigating the impact of disruptions, particularly during critical moments such as May 3.

However, a thorough evaluation of how redundancy is implemented, along with the supporting infrastructure, is necessary to reduce the likelihood of route server failure. Such an evaluation would help ensure that potential failures do not propagate to other route servers, thereby preserving network stability. 
The Internet Exchange infrastructure comprises geographically diverse interconnection points (PIX) where customers connect. This structure is linked to a central node. Two route servers exist, one in the central node and one in another PIX.

\textbf{Takeaway:} The IX infrastructure was largely unaffected; however, having the central point redundancy in the exact physical location raises concerns. It is necessary to evaluate how redundancy is implemented and assess the supporting infrastructure to reduce the risk of failure in the Route Servers and the central structure.

\section{SIMET Data: the End-User View}
\label{sec:simet}


SIMET, a Portuguese acronym for Traffic Measurement System~\cite{simet}, has measured Internet quality in Brazil since 2005. Operated by NIC.br on behalf of the Brazilian Internet Steering Committee, the system enables quality tests through two methods: (1) on-demand tests used by end-users via a website or mobile application and (2) periodic tests conducted by software installed on personal computers, often in government-operated schools~\cite{simet_escolas} and healthcare units~\cite{simet_saude}. This software is installed in 5,171 public schools across Rio Grande do Sul (vantage points), performing measurements every four hours. End-users within the state perform multiple on-demand measurements daily from over 100 different devices (vantage points).\reviewfix{Rephrased for clarity and conciseness}

In our first analysis, we compared the timeline of on-demand measurements from end users and schools to understand Internet activity in the region before, during, and after the event. During the analyzed period, we observed a notable increase in on-demand activity among end users. Specifically, 424 users conducted 1,085 tests between April 8 and April 12 (before the events), 567 users conducted 1,572 measurements between May 6 and May 10 (critical phase), and 474 users conducted 1,268 measurements between June 3 and June 7 (post-event) in Rio Grande do Sul. \reviewfix{B4}

\begin{figure}[htb!]
	\centering	
	\includegraphics[width=\textwidth]{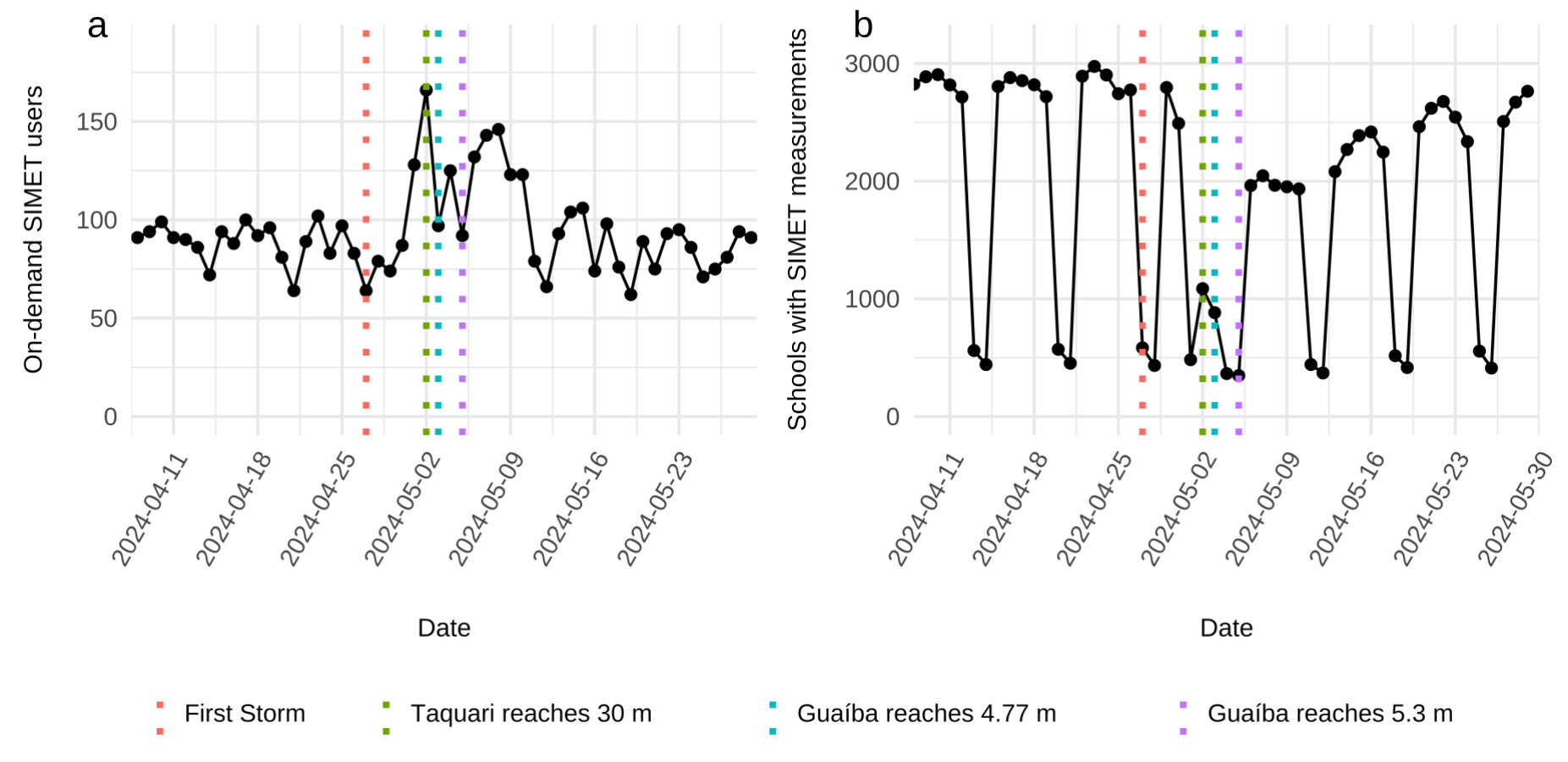}
	\caption{Aggregate view of SIMET vantage points for (a) end-users and (b) schools in Rio Grande do Sul state, covering the baseline (April 8-26), the critical phase (May 1-10), and the following period (from May 23 onward). Weekend activity is consistently lower.\protect\reviewfix{B4}}
	\label{fig:simet-TS}
\end{figure}

\autoref{fig:simet-TS}a illustrates the number of distinct end-user vantage points that performed measurements during this period. 
Historical data from the operator suggests that end users typically measure Internet quality when they experience or perceive poor connectivity. This behavior is evidenced by a 33\% increase in users and a 45\% increase in measurements during the critical phase. Notably, the number of users conducting measurements peaked on the day the Taquari River reached 30 meters, representing a 63\% increase compared to the highest measurement volume during the baseline period (April 8–26). Following this peak, the number of measurements gradually returned to typical levels. \reviewfix{B4}

It is important to note that this peak in activity was primarily observed in less affected cities. In contrast, the most affected areas experienced a decline in the number of users conducting measurements. This decline may indicate a broader reduction in users' ability to measure Internet quality during the most critical phases of the event. For additional details, refer to Appendix. \reviewfix{B4}


The observed increase in end-user activity on working days coincided with a decrease in school measurements during the event (\autoref{fig:simet-TS}b). Between April 8–12, 4,149 schools measured Internet quality at least once (around 3,000 per active day), compared to 3,182 schools from May 6–10. Recovery began on May 13, with 3,800 schools measuring Internet quality between May 27–31. This decline is attributed to school closures during the event, resulting in computers being turned off.


In our second analysis, we evaluate the spatial distribution of the end-users' and schools' measurements before and during the event. The overall increase in SIMET end-user measurements was not observed in the entire state. On-demand measurements increased in less affected cities but decreased in most affected ones (\textit{i.e.}, Porto Alegre, Canoas, Eldorado Do Sul, Gua\'{i}ba and Lajeado) during the event (\autoref{table:summ_measurements_period}). 

\begin{table}[htb!]
\centering
\small
\caption{Number of on-demand measurements for the most and least affected cities, along with the percentage of measurements for each period (before, during, and after the weather events of May 2024). The most affected cities are Porto Alegre, Canoas, Eldorado do Sul, Gua\'{i}ba, and Lajeado.}
\centering
\begin{tabular}[t]{rrrr}
\hline
Time period & Most affected cities & Other cities\\
\hline
08 to 12-April-2024 & 230 (21.2\%) & 855 (78.8\%)\\
\hline
06 to 10-May-2024 & 151 (9.6\%) & 1421 (90.4\%)\\
\hline
03 to 07-June-2024 & 305 (24.1\%) & 963 (75.9\%)\\
\hline
\end{tabular}
\label{table:summ_measurements_period}
\end{table}

\begin{figure}[htb!]
 	\centering
 	
 	\begin{subfigure}[b]{1\textwidth}
 		\centering
 		\includegraphics[width=.75\textwidth]{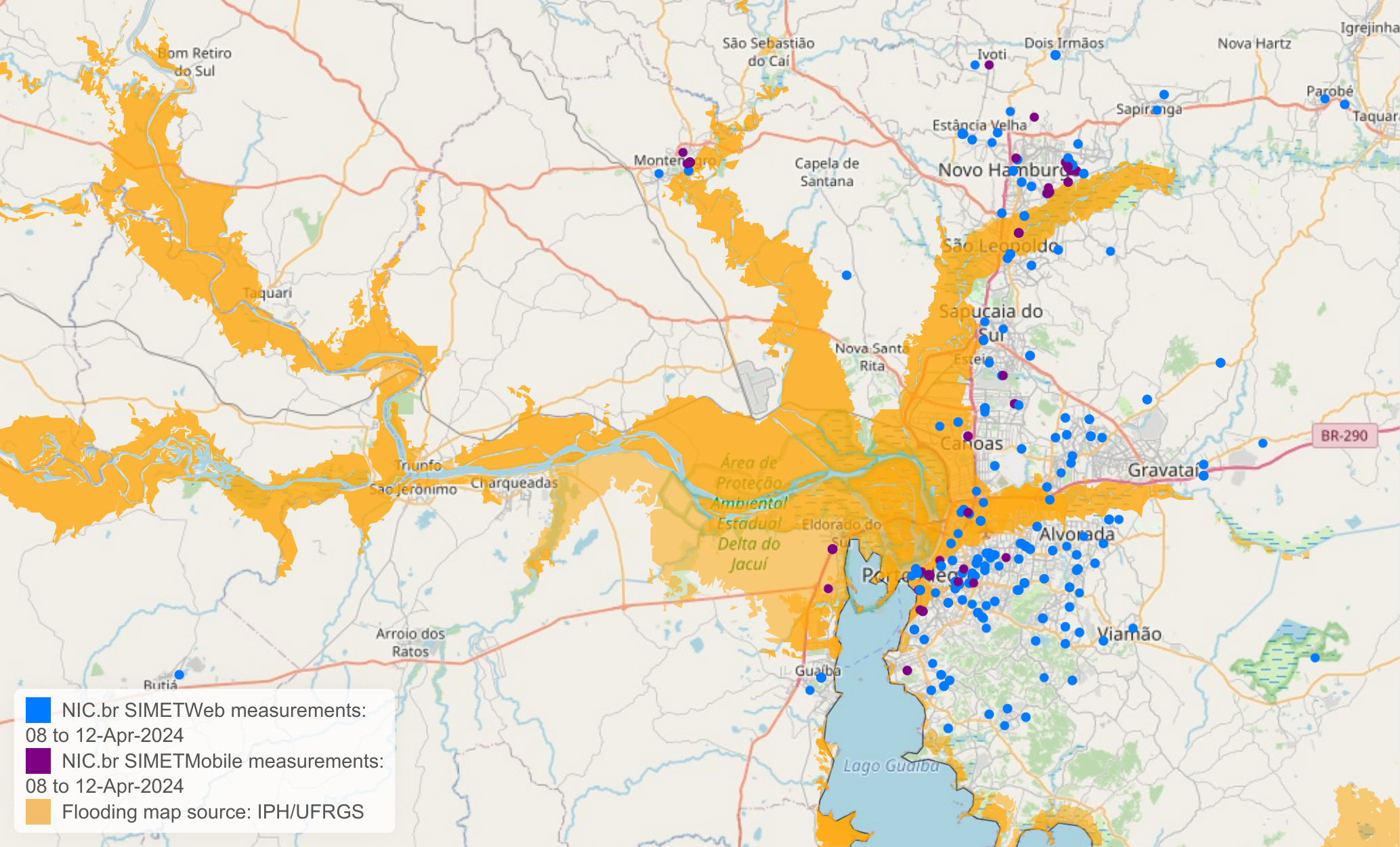}
 		\caption{Period before the event (08 to 12 April 2024).}
 	\end{subfigure}
 	\hfill

 	\begin{subfigure}[b]{1\textwidth}
 		\centering
 		\includegraphics[width=.75\textwidth]{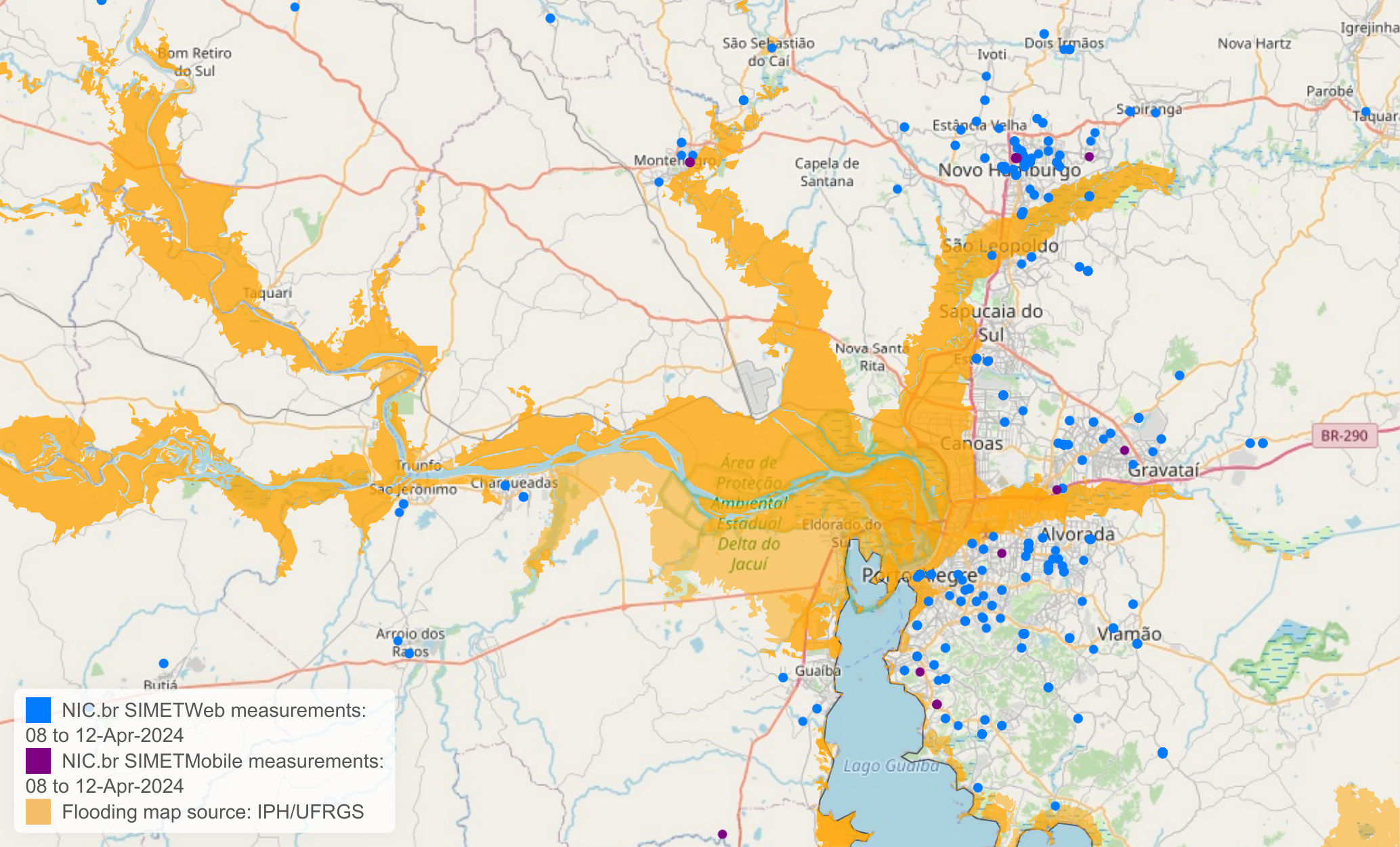}
 		\caption{Period during the event (6 to 10 May 2024).}
  		\label{fig:simet-any-after}
 	\end{subfigure}
  	\caption{End-users on-demand measurements received from Porto Alegre metropolitan area before and during the events in May 2024. Blue dots represent web measurements; purple dots represent measurements via mobile app. }
 	\label{fig:simet-comparison}
 \end{figure}

It is possible to observe a vacuum of measurements in the flooded area during the weeks following the flood (\autoref{fig:simet-any-after}). Although this decrease is not expected according to the idea that users measure the Internet quality when they are not satisfied, it can be explained by both a complete lack of Internet access, preventing users from being able to complete the measurement, or due to rearranging priorities (\textit{i.e.}, people were using less the Internet because they were worried about other things). Further investigations should be able to determine which of these factors, if any, was the most important. 

 For the less affected cities, the expected increase indicates a perception of worsened Internet quality followed by a recovery after the event. In \autoref{tab:simet_statistics}, we computed the quality measurements for end-users in the most affected cities. As mentioned, the median bandwidth for fixed and mobile Internet access diminished, but the mean round-trip-time (RTT) remained constant. A deeper investigation is needed to isolate areas inside each city. It is worth remembering that cities were divided between the flooding zones, where people lost their homes, and other parts, where life was close to normal.

 \begin{table}[htb!]
 	\centering
   		\caption{Statistics on SIMET measurements,  bandwidth, and RTT for fixed and mobile networks.}
 	\footnotesize
 	\begin{tabular}{|c|>{\centering\arraybackslash}p{1.8cm}|>{\centering\arraybackslash}p{1.8cm}|>{\centering\arraybackslash}p{1.8cm}|>{\centering\arraybackslash}p{1.5cm}|>{\centering\arraybackslash}p{1.5cm}|}

 			\hline
 			Time Period & Percentage of measurements from the total & Fixed Median bandwidth (Mbps) & Mobile Median bandwidth (Mbps) & Fixed Median RTT (ms) & Mobile Median RTT (ms) \\ \hline
 			08 to 12-April-2024 & 33.5\% & 94.4 & 119.5 & 31.9 & 28.6\\ \hline
 			06 to 10-May-2024 & 22.0\% & 92.7 & 101.2 & 31.3 & 23.7\\ \hline
 			03 to 07-June-2024 & 44.5\% & 92.0 & 69.5 & 33.4 & 36.7\\ \hline
 		\end{tabular}
 		\label{tab:simet_statistics}
 	\end{table}

\begin{figure}[htb!]
 	\centering
 	
 	\begin{subfigure}[b]{1\textwidth}
 		\centering
 		\includegraphics[width=.75\textwidth]{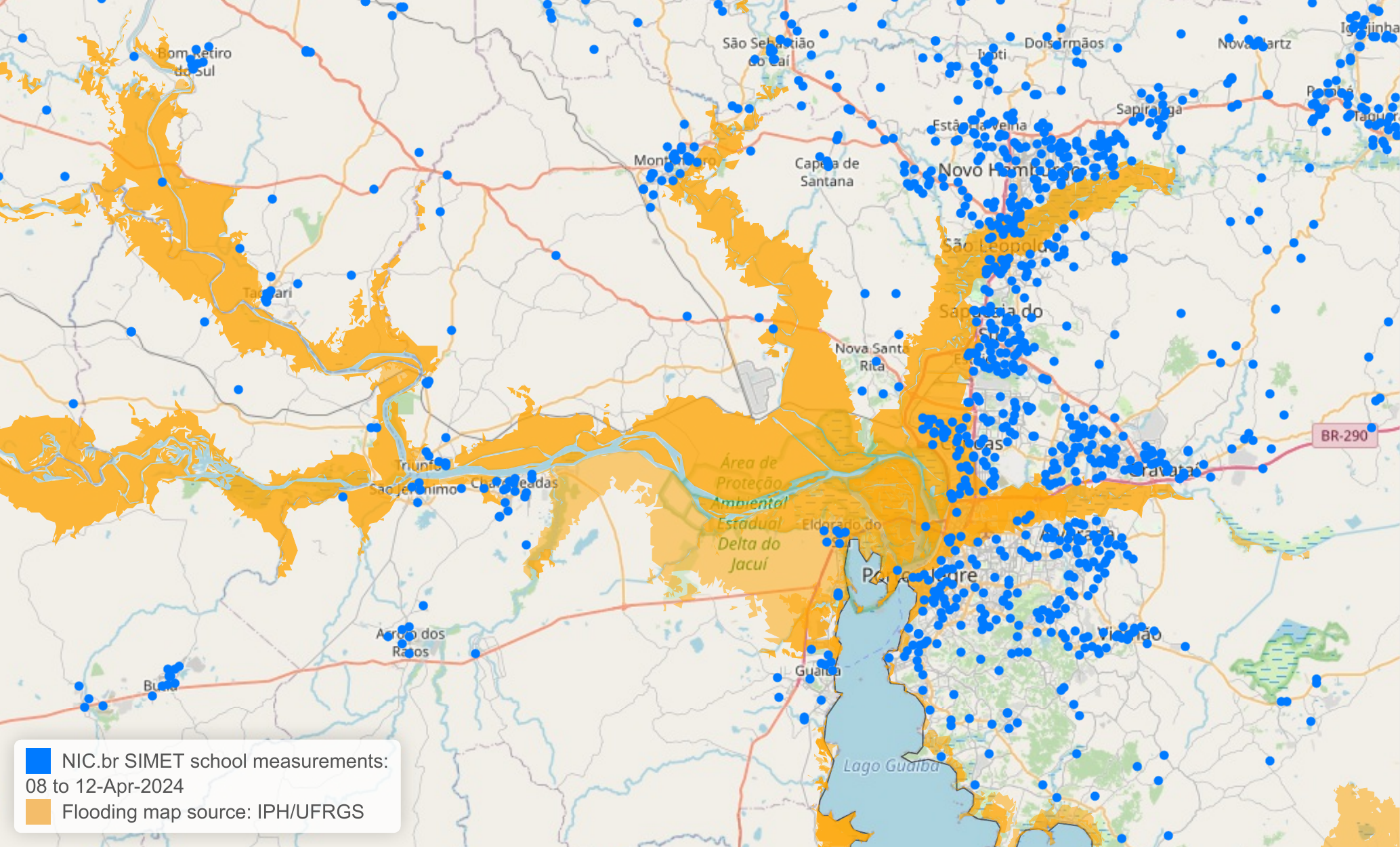}
		\caption{Period before the event (8 to 12 April 2024). }
 		\label{fig:simet-schools-before}
 	\end{subfigure}
 	\hfill
 	
 	\begin{subfigure}[b]{1\textwidth}
 		\centering
 		\includegraphics[width=.75 \textwidth]{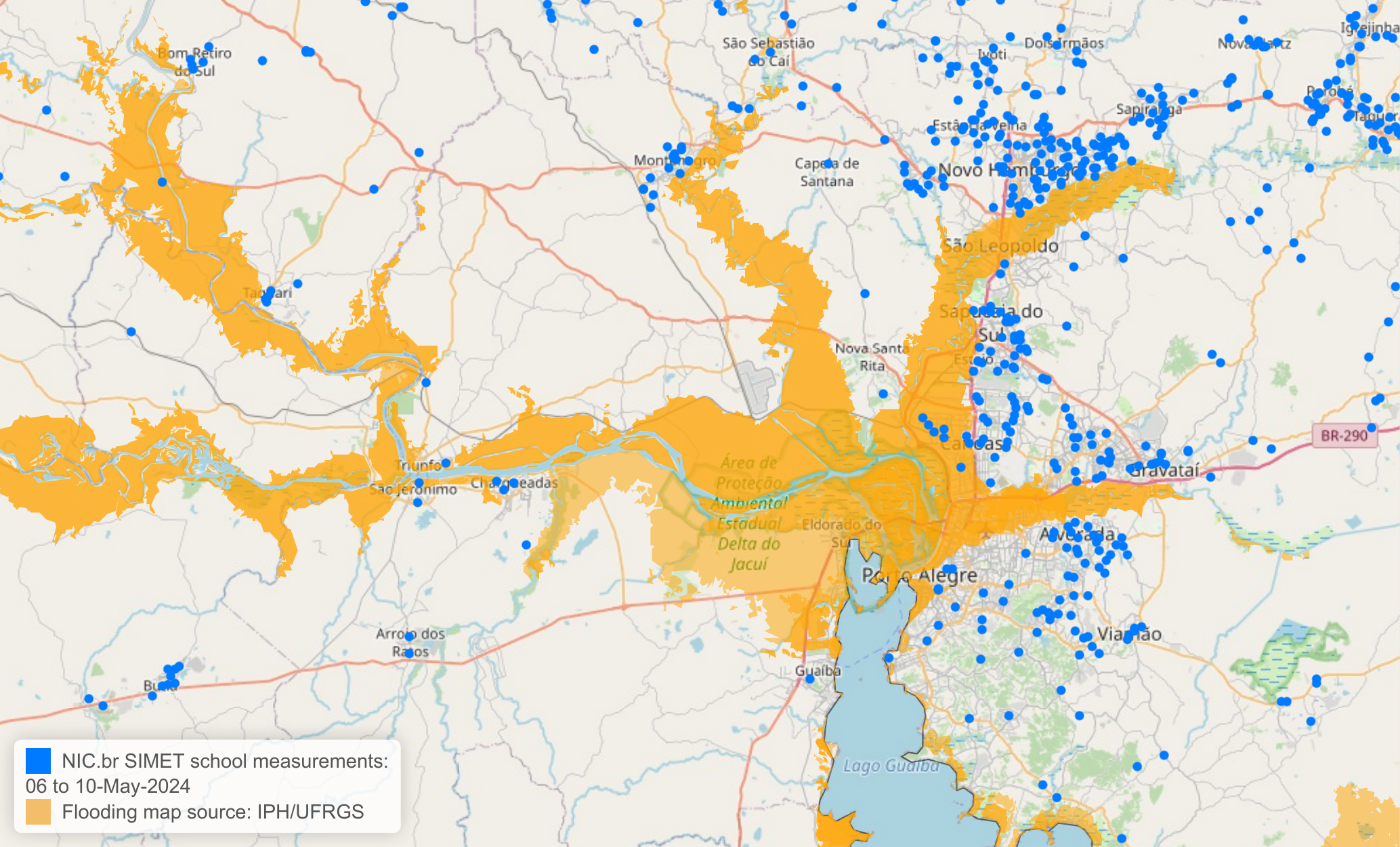}
 		\caption{Period during the event (6 to 10 May 2024).}
 		\label{fig:simet-schools-after}
 	\end{subfigure}
 	\caption{Measurements from public schools in 27 municipalities near the Porto Alegre metropolitan area with at least one Internet quality measurement recorded within a five-day period, before  and during the events of May 2024.}
 	\label{fig:simet-schools-comparison}
 \end{figure}

As expected, the difference in the number of schools measuring Internet quality is not homogeneously distributed in space. The most affected cities, usually representing 6.1\% of total measuring schools, suffered a greater decrease, representing 2.3\% and 4.4\% of measuring schools during and after the event. There was a decrease of 71\% of schools measuring in the most affected cities between April and the beginning of May compared to only 23\% in less affected cities. This decrease was observed in the flooded and immediate areas around it (\autoref{fig:simet-schools-comparison}). It is worth mentioning that among the six schools that sent measurements during the event and are located in a flooded area (\autoref{fig:simet-schools-after}), five changed the ASN and IP addresses during the event. These schools' measurements went through 5, 4, 2, 1, and 1 IP addresses and 4, 2, 2, 1, and 1 different AS, respectively. This suggests that the computers with the software installed were relocated and subsequently activated in these new locations, enabling measurements during this period. We visited five schools in June, which remained inoperative after the event, and confirmed that some computers had been moved to another place. \reviewfix{B5,B6}

\textbf{Takeaway: } It was possible to observe the effect of the flood on perceived end-user Internet experience along with its recovery for less affected cities considering both on-demand and school measurements. By combining these two types of measurements, it may be possible to detect critical failures in similar events.
\section{Post-Flooding Impact Assessment and Recovery Efforts}
\label{recovery}

Five months after the event, numerous institutions are still recovering their networks and data centers. Internet connection prices have remained stable, and providers are actively replacing home equipment that was damaged or lost during the floods at no additional cost to customers. Since the onset of the disaster, the ISP Provider Association has been actively involved in collecting donations and redistributing resources to various small ISPs to mitigate their losses and support the rebuilding of their networks. Several key indicators have been provided to guide these efforts~\cite{Internesul2024indicadores}. The provider association estimates a 36-month timeline to restore ISP operations to pre-event conditions, with total losses projected at R\$ 2 billion (US\$ 380 million).

Several institutions that previously operated their data centers have opted to migrate their operations to cloud-based solutions or relocate to unaffected data centers in other cities or states. The recent event has likely prompted one of the primary data center providers in the region to announce plans for constructing a new facility in an unaffected area. The proposed site is 34 kilometers from the existing data center and is situated near significant power lines~\cite{scala2024eldorado}. However, this presumed motivation has yet to be officially confirmed.

Of the 4,149 schools for which regular measurements were taken in April 2024, 181 (4.36\%) remain non-operational since the event. Some of these schools reported total losses of ICT equipment, with recovery contingent upon government funding to replace the lost assets.


\footnotetext[2]{Those ASes with a registrant (owner) address in Rio Grande do Sul, based on an AS list provided by NIC.br.}

\begin{figure}[htbp]
    \centering
    \includegraphics[width=\linewidth]{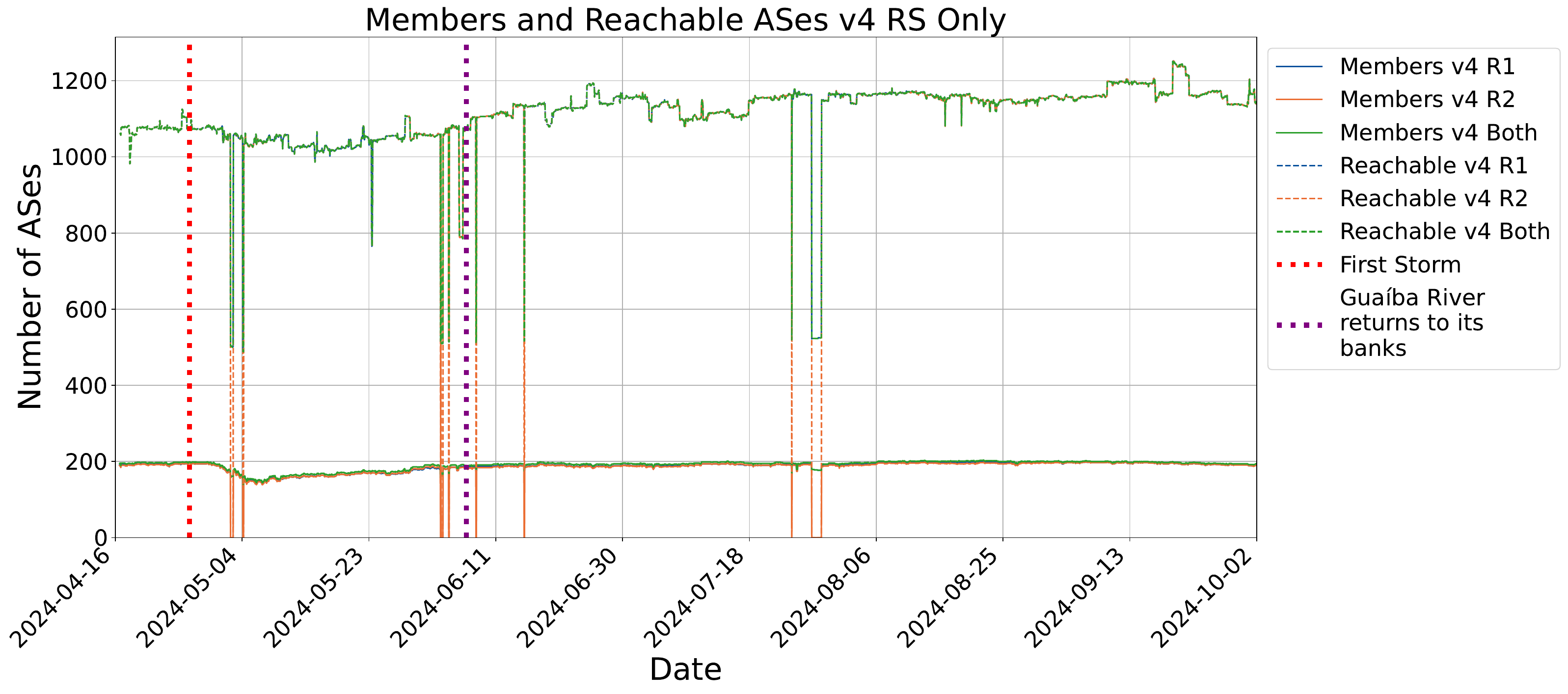}
    \caption{Reachable ASes from the Rio Grande do Sul Internet Exchange Point.}
    \label{fig:recuperacao_membros}
\end{figure}

The local Internet Exchange Point (IXP) has not yet recovered to its pre-event traffic volume, remaining approximately 23\% below prior levels. Despite this, all AS members have resumed active status. The number of ASes geolocated within the state\footnotemark[2] and reachable through the local IXP increased from 1,060 on April 16th to 1,141 on October 2nd, representing a 7\% growth and continuing the trend observed before the event (\autoref{fig:recuperacao_membros}). \reviewfix{B5}

IXP customers utilizing protected connectivity returned to normal parameters by June 2024 (take as reference \autoref{fig:fault-circuits}). However, customers relying on unprotected fiber continued to experience a significant number of incidents for several months after May 2024, as issues involving fiber networks persisted due to city reconstruction efforts (refer to \autoref{fig:metropoa-incidents} and Appendix). \reviewfix{C4}



\section{Related Work} 
\label{sec:related}

Infrastructure resilience during extreme weather events, particularly within the telecommunications sector, has been extensively studied. These studies include analyses of the economic impacts~\cite{frame2020climate,jahn2015economics}, explorations of resilience metrics and methodologies~\cite{ji2017resilience,reed2009methodology,tornatore2016survey,JI2024survey}, and investigations of the aftermath and recovery processes~\cite{cordova2021building,fcc2006report}. More recently, attention has shifted towards enhancing critical systems' resilience through machine learning approaches~\cite{Alkhaleel2024MachineLA,JI2024survey}.

Most of these studies examine short-duration events like hurricanes, earthquakes, and storms. To the best of our knowledge, the most relevant comparisons to our research, based on documented events, are the flooding in New Orleans following Hurricane Katrina~\cite{Kwasinski2006katrina,parker2009preventable,nanog36katrina} and in Puerto Rico following Hurricanes Irma and Maria~\cite{cordova2021building},  in 2005 and 2017, respectively.

Hurricane Katrina severely impacted the Gulf Coast's telecommunications infrastructure, particularly in New Orleans, where extensive flooding and high winds caused widespread damage. The storm overwhelmed much of the region’s communications infrastructure, including wireline networks and cellular systems. Over 70\% of cellular sites were rendered inoperable, and widespread flooding damaged underground cables and key communication nodes~\cite{clark2018cellKatrina}. In New Orleans, the failure of power supplies and fuel shortages for backup generators exacerbated the situation, leaving vast areas without communication for an extended period.
Major failures were linked to three key issues: (1) flooding that damaged infrastructure, (2) insufficient power and fuel for backup generators, and (3) the failure of redundant pathways, which severely affected network resilience. Additionally, limited coordination between the communications industry and government agencies further delayed response efforts, compounded by inadequate security measures for infrastructure and personnel.
A significant lesson learned from Katrina was the critical role of data accessibility in disaster response. The scarcity of publicly available data during the aftermath hindered recovery efforts, leading to initiatives such as the Climate Data Initiative to enhance open-data sharing~\cite{fcc2006report}. This issue persisted even a decade after the disaster~\cite{TenYearsNewOrleans}, prompting a governmental initiative to improve open-data access through efforts such as the Climate Data Initiative~\cite{ObamaClimateDataInitiative}.
The experiences from Katrina have informed current disaster preparedness and emphasized the importance of  reliable telecommunications during crises.

As discussed in \cite{cordova2021building}, \reviewfix{C4} following two hurricanes, Puerto Rico's wireline, wireless, and cable networks were mainly rendered inoperable. The wireline telecommunications infrastructure was particularly affected, as it predominantly relied on aerial fiber-optic cables for cellular backhaul. By being exposed to environmental elements, these cables were highly vulnerable to storm damage, primarily because many were supported by poorly maintained poles owned by the Puerto Rico Electric Power Authority (PREPA). The hurricanes caused extensive damage, with over 90\% of the private telecommunications infrastructure, including antennas and aerial cables, being affected. In the immediate aftermath, 95\% of cellular sites were out of service, and 48 of 78 municipalities completely lost cellular connectivity. As a result, citizens were without Internet access or the ability to reach emergency services, such as 911, for weeks or even months.
In response to the disaster, the Homeland Security Analysis Center proposed 33 courses of action (COAs) to improve the resilience of Puerto Rico’s telecommunications infrastructure. These include enhancing network robustness and redundancy, strengthening emergency communications, implementing regulatory reforms, and establishing funding mechanisms. Some proposals are notably innovative, such as involving local communities in the planning and implementation processes to ensure the infrastructure meets their specific needs, and fostering Public-Private Partnerships (PPP) to fund and deploy more resilient broadband infrastructure.

The severe weather events in the Rio Grande do Sul, New Orleans, and Puerto Rico share several similarities regarding their impacts on telecommunications infrastructure.
Both Puerto Rico and the Rio Grande do Sul heavily relied on aerial fiber-optic cables, which led to significant damage during hurricanes and floods. Similarly, New Orleans saw widespread damage to underground cables (Cooper) caused by flooding during Hurricane Katrina, and in Rio Grande do Sul, it was caused by landside and bridge collapse. 
Power and fuel shortages for generators were a common issue, and there were also reports of portable generators and other equipment being stolen.
Rio Grande do Sul faced similar power-related challenges, with critical data centers shutting down due to floodwaters, further disrupting services. 
The loss of connectivity was also a significant outcome across all cases. The number of outages in cellular service was similar, with around 50\% of cities being affected in Rio Grande do Sul~\cite{anatel2024celular}. 

Despite these similarities, there are notable differences between the three cases. In New Orleans, the damage primarily affected underground copper cables and communication nodes, whereas in Porto Alegre, although the underground fiber cables were inaccessible, no damage was reported, facilitating the recovery process for end-users broadband clients. The duration and extent of the impact also varied significantly. In Puerto Rico, connectivity losses persisted for months in some regions, reflecting a larger scale of infrastructure destruction. In contrast, in Rio Grande do Sul, the recovery process, though extensive, was already underway during the event. This rapid response may be attributed to the structure of the fiber and broadband market in Brazil. Large operators account for approximately 45\% of broadband users and 35\% of the optical fiber infrastructure, with the remaining market share distributed among thousands of smaller operators. These smaller operators collaborated by donating and sharing equipment, personnel, and fiber resources, which facilitated recovery efforts despite disruptions in transportation and supply chains. 

Another key distinction is that when the New Orleans Central Exchange Office (CEO) was flooded, a major failure occurred in the phone system, whereas in Rio Grande do Sul, critical operations were shifted to a secondary data center located on higher ground, minimizing the impact of the disaster. Regarding Internet access, in the New Orleans region, connectivity to the rest of the country experienced minor interruptions, with less than 10\% of ASes losing connectivity, while in Rio Grande do Sul, this figure exceeded 53\% in one day.

The comparison of telecommunications infrastructure impacts in New Orleans, Puerto Rico, and Rio Grande do Sul highlights common challenges, such as widespread connectivity loss, infrastructure damage, and power shortages. However, significant differences emerged in the scale of damage and recovery timelines. While Puerto Rico experienced prolonged outages due to extensive destruction of aerial infrastructure, Rio Grande do Sul’s recovery was more rapid, aided by market structure, collaborative efforts among smaller operators, and less damage in poles. Conversely, Rio Grande do Sul had a more significant impact on data centers in the area.

\section{Lessons Learned}
\label{sec:lessons}


The extreme weather event that struck Rio Grande do Sul in May 2024 is a significant case study for understanding the resilience of telecommunications and information technology infrastructure. The extensive damage caused by severe rainfall and flooding highlighted some lessons for improving and enhancing ICT operations. We list the main ones here:



\textbf{The Importance of Resilience in Fiber Infrastructure:} Underground fiber networks have demonstrated greater resilience than aerial fibers during extreme weather events. However, their effectiveness depends on  being \textit{``consistent''} before incidents such as flooding. A resilient city should prioritize the placement of primary optical paths underground and establish contingency plans for alternative interconnection routes in emergencies.

\textbf{Proactive Risk Management:} Robust risk analysis in data center standards is essential. The NBR ISO/IEC data center standards should include comprehensive business risk analyses covering all known terrain situations, not just a  window time of 5-10 years. This approach helps develop effective mitigation plans and ensures data centers are better prepared for such events.

\textbf{Community and Collaborative Efforts:} Community efforts and collaborations between ISPs, data centers, and government agencies were crucial in restoring services. Sharing equipment, cables, data center space, and resources greatly aided recovery efforts. Universities and academic infrastructure played a key role in supporting civil defense and other essential services, likely due to their tradition of collaboration with others.

\textbf{Redundancy and Diversification:} Redundant systems and diversified infrastructure routes are critical. The resilience demonstrated by the Rio Grande do Sul IXP through its redundant route servers minimized the impact during critical moments. ISPs and data centers should diversify paths and physical locations to prevent simultaneous failures during  disasters.

\textbf{Emergency Preparedness and Response:} Immediate response measures, such as opening mobile networks for free-roaming and providing emergency access to roads for ISPs, were vital in maintaining communication. Planning for fuel transport and generator maintenance is crucial for data centers during prolonged power outages.

\textbf{Impact of Natural and Human Factors:} The flooding in the metropolitan area was intensified by natural factors (heavy rainfall and strong winds) and human factors (lack of maintenance and updates to flood protection systems). Addressing the human factors is essential to mitigate future risks.

\textbf{Adaptation and Long-term Recovery:} Long-term recovery should focus on strengthening infrastructure to withstand future climate events. Shifting to cloud solutions and relocating critical services to safer areas are key strategies. Reconstruction, potentially taking up to 36 months, should prioritize resilience.

\textbf{Public Awareness and Communication:} Effective communication systems are vital for coordinating rescues, informing the public, and maintaining order during disasters. Raising awareness about the importance of  telecommunications infrastructure can help in future impacts.

These lessons highlight the long-term need for comprehensive planning, robust infrastructure, community collaboration, and proactive risk management to achieve resilience against extreme weather events and ensure the continuity of ICT services.
Internet providers are rebuilding their infrastructure in the short term by exchanging equipment and fiber cables and seeking donations. Several companies have contributed thousands of dollars. Reestablishing customers' Internet access and equipment is expensive and time-consuming. The providers association estimates it will take three years to restore the ISPs' infrastructure in the area fully.

\section{Open Challenges and Future Work}
\label{challenges}

\reviewfix{A8,B6}
Despite significant progress in understanding, modeling, and addressing the resilience of ICT and power infrastructure during extreme weather events, several challenges still need to be discussed for further research.

\textbf{Modeling Resilience at Scale}: One key challenge is developing comprehensive models that quantify resilience across large-scale, dynamic, and dependent networks. These models must account for power, weather conditions, localized failures, and other variables. Newresearch studies are trying to improve both the accuracy and scalability of these resilience models \cite{JI2024survey}.

\textbf{Data Availability and Analysis}: Another critical challenge is obtaining detailed, large-scale data  for assessing resilience. 
Data on power distribution grids, ICT failures, and weather events are often fragmented and owned by various entities and private companies.
Collaboration between these entities, policymakers, and researchers is essential to
access this data and conduct the necessary analysis. The first step is to provide
open or easily accessible data \cite{ObamaClimateDataInitiative}

\textbf{Developing Resilience Metrics}: Existing reliability metrics, such as SAIFI (System Average Interruption Frequency Index) and SAIDI (System Average Interruption Duration Index) from the electrical power system, or MTTR (Mean Time to Repair) and MTBF (Mean Time Between Failures) from the ICT domain, are insufficient for characterizing resilience during severe weather events, as they were primarily designed for daily operations. There is a need to develop resilience metrics that incorporate external factors, such as weather variables and service restoration timelines. Such metrics require a multidisciplinary approach, integrating infrastructure, service management, and meteorology.

\textbf{Leveraging Machine Learning and Data-Driven Methods}: Integrating machine learning with spatiotemporal random processes to predict failures and optimize recovery is a promising yet challenging area. For example, using Phasor Measurement Units (PMUs) and ICT infrastructure data could improve failure detection and service restoration in smart cities. Still, these models account for complex, large-scale interdependencies.

\textbf{Improving Measurement Infrastructure in the Region:} Another avenue for exploration is the use of traceroutes passing through the affected area. The limited probing infrastructure in the region, such as RIPE Atlas~\cite{ripe2015atlas}, complicates comparisons with other events, such as those involving IXPs \cite{bertholdo2021forecasting}. \reviewfix{B3}

\textbf{Deeper Analysis:} Our initial analysis provides valuable insights into the local impact of the event, but several unanswered questions require deeper investigation and complementary data for a broader perspective. For instance, understanding why the IXP’s traffic volume has not normalized six months after the event may involve analyzing the status of ASes before and after the event using additional datasets, such as from  IODA (Internet Outage Detection and Analysis)~\cite{ioda} and Trinocular (ICMP probes to detect outages)~\cite{quan2013trinocular}\cite{ANTOutageMap}. \reviewfix {C3}





Addressing these challenges will require interdisciplinary collaboration and the development of novel methodologies that integrate data-driven approaches, advanced modeling techniques, measurement infrastructure, and innovative resilience metrics. The results could significantly improve our ability to prepare for and respond to extreme weather events, ensuring more resilient infrastructure systems in the future.

\section{Availability of datasets and additional materials}
\label{datasets}
In this study, we are cataloging data for future analyses and simulations. This data includes event timelines, hydrological causes, maps, data center conditions, and Internet measurements collected from end-users, public schools, and health units through the Simet system~\cite{simet}. Additionally, we have gathered Internet routing and traffic data from the Porto Alegre Internet Exchange Point (IXP) (IX.br) and a mobile operator, documented fiber cuts and repairs across two metropolitan networks, and collected information on long-distance circuit outages. This dataset is supplemented with data from operator management systems, private sector contributions, and official reports.

We are progressively making intermediate datasets available to the community, as detailed in~\cite{github2024furg}, subject to obtaining the  authorizations from all  stakeholders. 
All information used in this study is accurate and reflects the most recent data available as of September 2024.

\section{Ethical considerations}

This study utilizes anonymized and non-anonymized datasets from public sources and private entities, all of which have been published with the necessary consent. No personal or sensitive data were involved in the research. Data integrity is ensured through a collaborative effort between academic institutions and private operators, with transparency and interdisciplinary partnerships forming the foundation of trust. The primary objective of this study is to educate, assess, and facilitate learning through a neutral, non-judgmental approach.

\section{Conclusion}

In this paper, we have contributed by documenting the sequence of extreme weather events in May 2024 in Rio Grande do Sul, Brazil, specifically focusing on the resilience and recovery of Information and Communication Technology (ICT) infrastructure. The study assesses the impact of these events on various critical infrastructure components, including fiber optic networks, data centers, and the regional Internet Exchange Point (IXP), while also providing insights from an end-user perspective during the critical period.
Our findings highlight the vulnerability of aerial fiber networks to climate-related disruptions and underscore the need for proactive measures to fully utilize the relative resilience of underground networks. They corroborate similar observations from events in Puerto Rico and New Orleans. Furthermore, we provide critical insights into the performance of affected data centers, which may inform discussions on risk analysis standards, such as  ISO/IEC 22237.
The study demonstrates a direct correlation between climate events in the region and increasingly frequent disruptions to aerial fiber infrastructure. We have compiled data from multiple sources with the aim of making it openly accessible for future research, particularly for machine learning studies on the resilience of critical infrastructure.
Ultimately, our study reinforces the necessity of improved disaster preparedness, including enhancements in redundancy, risk management, and infrastructure planning, to mitigate the impact of future extreme weather events on ICT infrastructure.

\newpage
\begin{subappendices}
\renewcommand{\thesection}{}%
	
\section{Additional Information and Graphs} 
\label{app:graph}
	
\reviewfix{B2, B4, C2, S1, D5}
	
\begin{figure}[H]
		\centering
		\includegraphics[width=.85\textwidth]{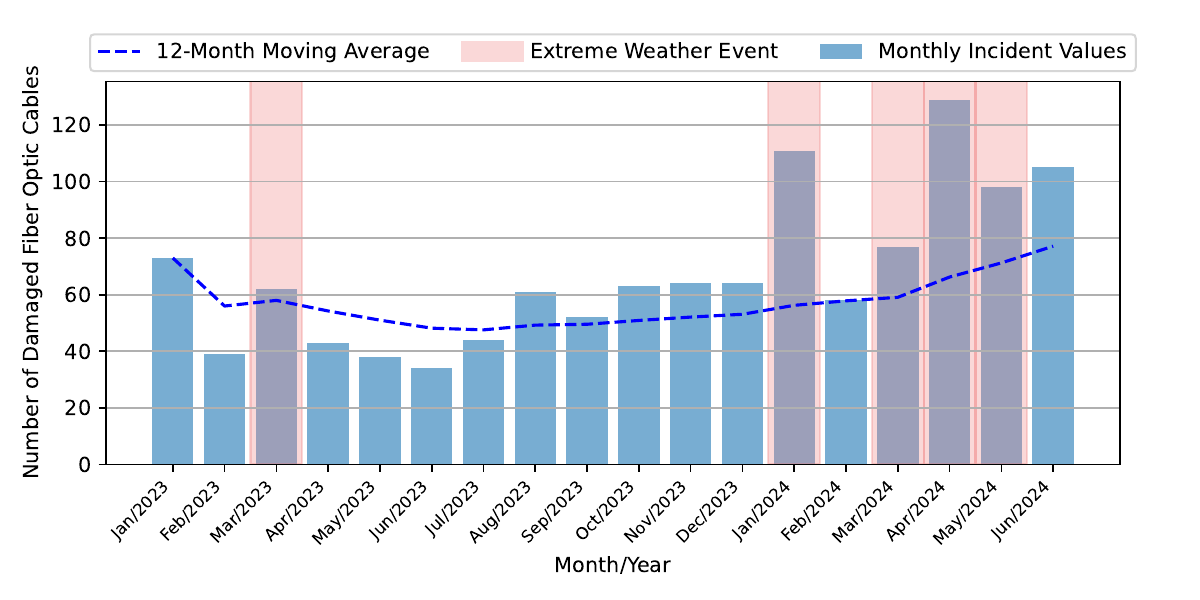}
	\caption{Incidents affecting fiber paths in a 500 km ISP backbone in the region.}
			\label{fig:sebratel-incidents}
		\end{figure}

\textbf{Fiber damage:} the Metro datasets in \cite{github2024furg} provide detailed information on the dates, causes, and locations of incidents affecting two fiber backbones. \autoref{fig:sebratel-incidents} presents data on incidents reported by a commercial (ISP) fiber backbone, showing trends consistent with those discussed in Section \ref{sec:aerial}.
	
\begin{figure}[H]
		\centering
		\includegraphics[width=.85\textwidth]{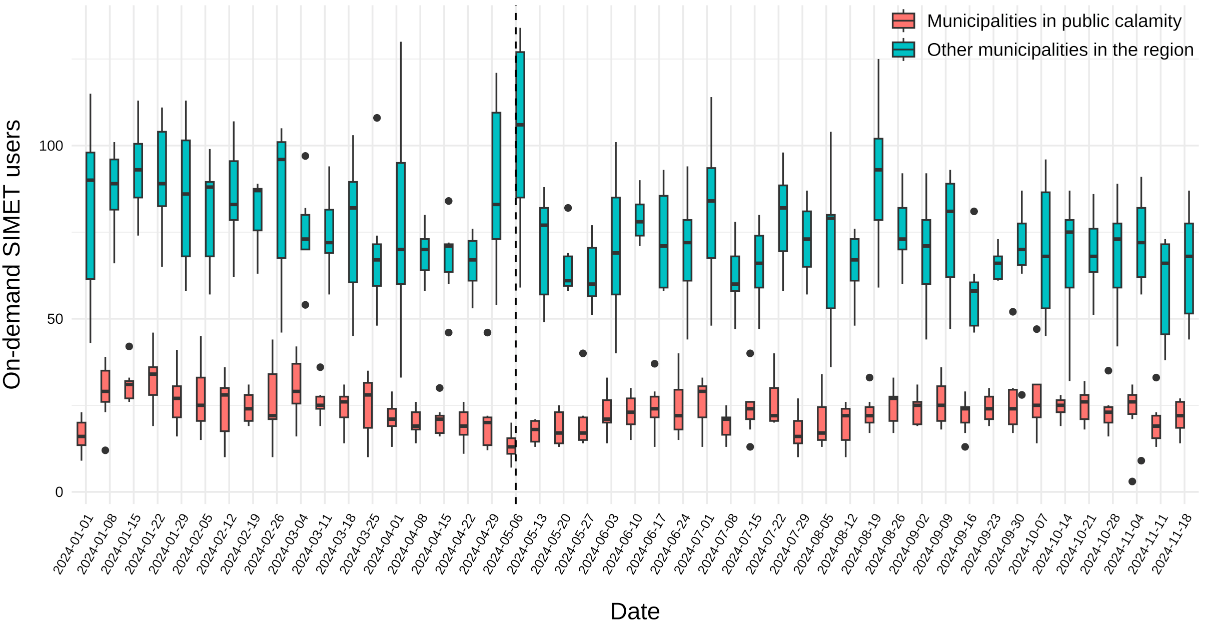}
		\caption{Long-term weekly analysis of median and percentile metrics per unique end-user vantage point from January to November 2024. The dashed line indicates the worst week of the flooding in May 2024. }
		\label{fig:simet-median}
	\end{figure}

	\textbf{SIMET:} the dataset provides comprehensive evaluations of Internet quality, measuring key performance metrics such as download and upload speeds, latency, jitter, and packet loss (via TWAMP). It also collects contextual data, including timestamps, IP protocol version (IPv4 or IPv6), IPv6 connectivity availability, NAT or CGNAT presence, and, with user consent, geolocation information. In \autoref{fig:simet-median}, the time window is extended to enhance baseline identification and to present a breakdown by city groups within Rio Grande do Sul.
	
\end{subappendices}

\bibliographystyle{plain}
\bibliography{refs}

\begin{thebibliography}{10}

\bibitem{Alkhaleel2024MachineLA}
Basem~A. Alkhaleel.
\newblock Machine learning applications in the resilience of interdependent
  critical infrastructure systems - a systematic literature review.
\newblock {\em Int. J. Crit. Infrastructure Prot.}, 44:100646, 2024.

\bibitem{anatel2024celular}
Anatel.
\newblock Anatel - recuperação das redes.
\newblock
  \url{https://informacoes.anatel.gov.br/paineis/utilidade-publica/recuperacao-das-redes}.
\newblock (Accessed on 10/08/2024).

\bibitem{github2024furg}
Leandro Bertholdo and Renan Barreto.
\newblock Rsfloodsdatasets.
\newblock \url{https://github.com/systems-furg/RSFloodsDataset/}, September
  2024.

\bibitem{bertholdo2021forecasting}
Leandro~Marcio Bertholdo, Jo{\~a}o~M Ceron, Lisandro~Zambenedetti Granville,
  and Roland van Rijswijk-Deij.
\newblock Forecasting the impact of ixp outages using anycast.
\newblock In {\em TMA}, 2021.

\bibitem{brazil_heavy_rains2024bbc}
Vanessa Buschschlüter.
\newblock Brazil floods:hundreds of rio grande do sul towns under water.
\newblock
  \footnotesize\url{https://www.bbc.com/news/world-latin-america-68968987}, May
  2024.

\bibitem{lei137070poa}
{CamaraPOA}.
\newblock {Camara de Vereadores de Porto Alegre Lei 13.402}.
\newblock
  \footnotesize\url{https://camarapoa.rs.gov.br/draco/processos/137070/Lei\_13402.pdf},
  May 2023.

\bibitem{simet}
{CEPTRO}.
\newblock Medições - iniciativas para analisar e melhorar a qualidade e
  velocidade da internet no brasil.
\newblock \footnotesize\url{https://medicoes.nic.br/}, Jun 2024.

\bibitem{roaming_celular}
{CNN}.
\newblock Operadoras de telefonia liberam sinal e oferecem internet no rs para
  facilitar comunicação.
\newblock
  \footnotesize\url{https://www.cnnbrasil.com.br/economia/macroeconomia/operadoras-de-telefonia-liberam-sinal-e-oferecem-internet-no-rs-para-facilitar-comunicacao/},
  May 2024.

\bibitem{temporal_mar2024cnn}
{CNN}.
\newblock Temporal e ventos de 90 km/h causam estragos em porto alegre e deixam
  bairros sem energia.
\newblock
  \footnotesize\url{https://www.cnnbrasil.com.br/nacional/temporal-causa-estragos-em-porto-alegre-e-deixa-bairros-sem-energia/},
  Mar 2024.

\bibitem{tjrs-offline2024conjur}
Conjur.
\newblock {TJ-RS} suspende prazos processuais e só analisa medidas urgentes.
\newblock
  \footnotesize\url{https://www.conjur.com.br/2024-mai-07/tj-rs-suspende-prazos-processuais-e-so-analisa-medidas-urgentes/},
  May 2024.

\bibitem{trf4-offline2024convergenciadigital}
{Convergencia Digital}.
\newblock {Justiça desliga data center no Rio Grande do Sul}.
\newblock
  \footnotesize\url{https://convergenciadigital.com.br/mercado/justica-desliga-data-center-no-rio-grande-do-sul/},
  May 2024.

\bibitem{cordova2021building}
Amado Cordova, Karlyn Stanley, Ajay Kochhar, Ryan Consaul, and Justin Hodiak.
\newblock {Building a Resilient Telecommunications Sector in Puerto Rico in the
  Aftermath of Hurricanes Irma and Maria}.
\newblock {\em Journal of Critical Infrastructure Policy}, 2(1):75--96, 2021.

\bibitem{procergs-offline2024cpovo}
{CPovo}.
\newblock Com sede alagada, {Procergs} desliga datacenter e sites do governo do
  rs ficam fora do ar temporariamente.
\newblock
  \footnotesize\url{https://www.correiodopovo.com.br/not%C3%ADcias/cidades/com-sede-alagada-procergs-desliga-datacenter-e-sites-do-governo-do-rs-ficam-fora-do-ar-temporariamente-1.1492060},
  May 2024.

\bibitem{internetsul_dnit}
{DNIT}.
\newblock Sei/dnit - 17714322 - ofício.
\newblock
  \footnotesize\url{https://internetsul.com.br/files/Oficio_17714322%20(DNIT).html},
  May 2024.

\bibitem{elea2offline2024}
Elea.
\newblock {Em meio às enchentes de Porto Alegre, Elea continua operação na
  cidade}.
\newblock
  \footnotesize\url{https://eleadatacenters.com/2024/05/07/em-meio-as-enchentes-de-porto-alegre-elea-digital-data-centers-continua-operacao-na-cidade/}.

\bibitem{fcc2006report}
FCC.
\newblock Report and recommendations of the independent panel reviewing the
  impact of hurricane katrina on communications networks.
\newblock \url{https://transition.fcc.gov/pshs/docs/advisory/hkip/karrp.pdf},
  Jun 2006.

\bibitem{fiergs2024industrias}
FIERGS.
\newblock Indústria segue contabilizando perdas com as inundações no rio
  grande do sul.
\newblock
  \footnotesize\url{https://www.fiergs.org.br/noticia/industria-segue-contabilizando-perdas-com-inundacoes-no-rio-grande-do-sul},
  May 2024.

\bibitem{frame2020climate}
David~J Frame, Suzanne~M Rosier, Ilan Noy, Luke~J Harrington, Trevor
  Carey-Smith, Sarah~N Sparrow, D{\'a}ith{\'\i}~A Stone, and Samuel~M Dean.
\newblock Climate change attribution and the economic costs of extreme weather
  events: a study on damages from extreme rainfall and drought.
\newblock {\em Climatic Change}, 162:781--797, 2020.

\bibitem{4thdistrito}
{G1RS}.
\newblock Câmara de porto alegre aprova incentivos ao 4º distrito.
\newblock
  \footnotesize\url{https://g1.globo.com/rs/rio-grande-do-sul/noticia/2022/08/17/camara-de-porto-alegre-aprova-projeto-de-incentivos-ao-4o-distrito.ghtml},
  Aug 2022.

\bibitem{temporal_mai2022globo}
G1RS.
\newblock {Tempestade Yakecan deixa mais de 200 mil clientes sem energia
  elétrica no RS}.
\newblock
  \footnotesize\url{https://g1.globo.com/rs/rio-grande-do-sul/noticia/2022/05/18/tempestade-yakecan-deixa-familias-sem-energia-eletrica-no-rs-alerta-para-ventos-fortes-seguem-nesta-quarta.ghtml},
  May 2022.

\bibitem{guaibaretornaleito}
{G1RS}.
\newblock Guaíba fica abaixo da cota de alerta e registra menor marca em mais
  de um mês em porto alegre.
\newblock
  \url{https://g1.globo.com/rs/rio-grande-do-sul/noticia/2024/06/07/nivel-guaiba-cota-de-alerta-porto-alegre.ghtml},
  Jun 2024.

\bibitem{abaixocota1jun}
{G1RS}.
\newblock Nível do guaíba fica abaixo da cota de inundação pela primeira
  vez em um mês em porto alegre.
\newblock
  \url{https://g1.globo.com/rs/rio-grande-do-sul/noticia/2024/06/01/nivel-do-guaiba-fica-abaixo-da-cota-de-inundacao-pela-primeira-vez-em-um-mes-em-porto-alegre.ghtml},
  Jun 2024.

\bibitem{guaibavoltasubir}
{G1RS}.
\newblock Nível no guaíba volta a subir, ultrapassa cota de inundação e
  porto alegre registra pontos de alagamento.
\newblock
  \url{https://g1.globo.com/rs/rio-grande-do-sul/noticia/2024/06/03/nivel-no-guaiba-volta-a-subir-ultrapassa-cota-de-inundacao-e-porto-alegre-registra-pontos-de-alagamento.ghtml},
  Jun 2024.

\bibitem{g1_baciaguaiba}
{G1RS}.
\newblock Temporais no {RS}: entenda como o relevo de porto alegre e as 'marés
  de tempestade' travam escoamento.
\newblock
  \footnotesize\url{https://g1.globo.com/meio-ambiente/noticia/2024/05/07/temporais-no-rs-entenda-como-o-relevo-de-porto-alegre-e-as-mares-de-tempestade-travam-escoamento.ghtml},
  May 2024.

\bibitem{cronologia2024g1}
{G1RS}.
\newblock Temporais no {RS}: veja cronologia de desastre que matou 75 pessoas.
\newblock
  \footnotesize\url{https://g1.globo.com/rs/rio-grande-do-sul/noticia/2024/05/05/temporais-no-rs-veja-cronologia-de-desastre.ghtml},
  May 2024.

\bibitem{ioda}
Gatech.
\newblock Ioda | internet outage detection and analysis.
\newblock \url{https://ioda.live}.

\bibitem{temporal_mar2023gzh}
GZH.
\newblock {Ao menos 52 mil pontos estão sem luz no RS em razão de temporais;
  Porto Alegre registra alagamentos em ruas}.
\newblock
  \footnotesize\url{https://gauchazh.clicrbs.com.br/geral/noticia/2023/03/ao-menos-52-mil-pontos-estao-sem-luz-no-rs-em-razao-de-temporais-porto-alegre-registra-alagamentos-em-ruas-clfudxn8b00250151yylijzxw.html},
  Mar 2023.

\bibitem{dmae_falha}
{GZH}.
\newblock Falhas de manutenção e de projeto ampliaram nível da inundação
  em porto alegre.
\newblock
  \url{https://gauchazh.clicrbs.com.br/porto-alegre/noticia/2024/05/falhas-de-manutencao-e-de-projeto-ampliaram-nivel-da-inundacao-em-porto-alegre-clw15ll0y00s6011hyc1p80ro.html},
  May 2024.

\bibitem{GZH2024onibus}
{GZH}.
\newblock Morte de morador desaparecido motivou protesto com queima de ônibus
  em porto alegre.
\newblock
  \footnotesize\url{https://gauchazh.clicrbs.com.br/seguranca/noticia/2024/05/morte-de-morador-desaparecido-motivou-protesto-com-queima-de-onibus-em-porto-alegre-clwf65q52004x014x9uqbg0aa.html},
  May 2024.

\bibitem{clark2018cellKatrina}
The~Sun Herald.
\newblock Will your cell service work if a hurricane rolls through the coast,
  and will it be enough?
\newblock
  \url{https://www.govtech.com/em/disaster/will-your-cell-service-work-if-a-hurricane-rolls-through-the-coast-and-will-it-be-enough.html},
  Jan 2018.

\bibitem{Internesul2024indicadores}
InternetSul.
\newblock {Associação dos Provedores de Serviços e Informações da
  Internet}.
\newblock \url{https://internetsul.com.br/ajuders}.

\bibitem{internetsul2023dutos}
InternetSul.
\newblock Reunião com a secretaria municipal para discutir os desafios e
  custos de dutos e cabos subterrâneos.
\newblock
  \url{https://internetsul.com.br/noticias/internetsul-se-reuniu-com-a-secretaria-municipal-de-parcerias},
  May 2023.

\bibitem{previsaoIPH}
IPH-UFRGS.
\newblock Previsões atualizadas de níveis d'água no guaíba – terça-feira
  04/06/24.
\newblock
  \footnotesize\url{https://www.ufrgs.br/iph/previsoes-atualizadas-de-niveis-dagua-no-guaiba-terca-feira-04-06-24-12h/},
  Jun 2024.

\bibitem{jahn2015economics}
Malte Jahn.
\newblock Economics of extreme weather events: Terminology and regional impact
  models.
\newblock {\em Weather and Climate Extremes}, 10:29--39, 2015.

\bibitem{JI2024survey}
Ankang Ji, Renfei He, Weiyi Chen, and Limao Zhang.
\newblock Computational methodologies for critical infrastructure resilience
  modeling: A review.
\newblock {\em Advanced Engineering Informatics}, 62:102663, 2024.

\bibitem{ji2017resilience}
Chuanyi Ji, Yun Wei, and H~Vincent Poor.
\newblock Resilience of energy infrastructure and services: Modeling, data
  analytics, and metrics.
\newblock {\em IEEE}, 105(7):1354--1366, 2017.

\bibitem{ufrgs2024cedep}
{Jornal da Universidade}.
\newblock O sistema de proteção contra inundações de porto alegre.
\newblock
  \url{http://www.ufrgs.br/ufrgs/noticias/o-sistema-de-protecao-contra-inundacoes-de-porto-alegre},
  Jun 2024.

\bibitem{Kwasinski2006katrina}
Alexis Kwasinski, Wayne~W. Weaver, Patrick~L. Chapman, and Philip~T. Krein.
\newblock Telecommunications power plant damage assessment caused by hurricane
  katrina - site survey and follow-up results.
\newblock In {\em Twenty-Eighth International Telecommunications Energy
  Conference}, pages 1--8, 2006.

\bibitem{metsul1000mm}
Metsul.
\newblock Chuva que levou às enchentes no rio grande do sul superou 1000 mm.
\newblock
  \footnotesize\url{https://metsul.com/chuva-que-levou-as-enchentes-no-rio-grande-do-sul-superou-1000-mm/},
  Jun 2024.

\bibitem{temporal_jan2024metsul}
Metsul.
\newblock Porto alegre tem noite de pavor com temporal violento e destrutivo.
\newblock
  \footnotesize\url{https://metsul.com/2024-01-17-temporal-porto-alegre-vendaval/},
  Jan 2024.

\bibitem{nanog36katrina}
{NANOG}.
\newblock {Hurricane Katrina Telecom Infrastructure Impacts, Solutions, and
  Opportunities}.
\newblock \url{https://www.youtube.com/watch?v=oel1jG9tOIM}, Feb 2006.

\bibitem{sara_project}
NIC.br.
\newblock Projeto sara.
\newblock \url{https://sara.nic.br/}, May 2024.

\bibitem{parker2009preventable}
Charles~F Parker, Eric~K Stern, Eric Paglia, and Christer Brown.
\newblock {Preventable catastrophe? The hurricane Katrina disaster revisited}.
\newblock {\em Journal of Contingencies and Crisis management}, 17(4):206--220,
  2009.

\bibitem{site_metropoa}
PoP-RS/RNP.
\newblock Rede metropoa.
\newblock \footnotesize\url{https://metropoa.tche.br/sobre/}, May 2024.

\bibitem{ANTOutageMap}
Yuri Pradkin, Kari Quan, and John Heidemann.
\newblock {ANT Outage World Map}.
\newblock \url{https://outage.ant.isi.edu/}, Dec 2024.

\bibitem{quan2013trinocular}
Lin Quan, John Heidemann, and Yuri Pradkin.
\newblock Trinocular: Understanding internet reliability through adaptive
  probing.
\newblock {\em ACM SIGCOMM Computer Communication Review}, 43(4):255--266,
  2013.

\bibitem{reed2009methodology}
Dorothy~A Reed, Kailash~C Kapur, and Richard~D Christie.
\newblock Methodology for assessing the resilience of networked infrastructure.
\newblock {\em IEEE Systems Journal}, 3(2):174--180, 2009.

\bibitem{procergs-offline2024baguete}
Mauricio Renner.
\newblock Procergs desliga data center.
\newblock
  \footnotesize\url{https://www.baguete.com.br/noticias/07/05/2024/procergs-desliga-data-center},
  May 2024.

\bibitem{richter2014peering}
Philipp Richter, Georgios Smaragdakis, Anja Feldmann, Nikolaos Chatzis, Jan
  Boettger, and Walter Willinger.
\newblock Peering at peerings: On the role of ixp route servers.
\newblock In {\em Proceedings of the 2014 Conference on Internet Measurement
  Conference}, pages 31--44, 2014.

\bibitem{defesacivil_municipios}
CASA MILITAR DEFESA~CIVIL RS.
\newblock Decreto amplia número de municípios em estado de calamidade e em
  situação de emergência - defesa civil do rio grande do sul.
\newblock
  \footnotesize\url{https://www.defesacivil.rs.gov.br/decreto-amplia-numero-de-municipios-em-estado-de-calamidade-e-em-situacao-de-emergencia},
  May 2024.

\bibitem{defesacivil_balanco}
Defesa~Civil RS.
\newblock Defesa civil atualiza balanço das enchentes no rs – 3/6, 9h -
  portal do estado do rio grande do sul.
\newblock
  \footnotesize\url{https://estado.rs.gov.br/defesa-civil-atualiza-balanco-das-enchentes-no-rs-3-6-9h},
  Jun 2024.

\bibitem{scala2024eldorado}
RS-Gov.
\newblock {Com investimento de R\$ 3 bilhões, Estado e Scala Data Centers
  assinam acordo para maior projeto de infraestrutura digital do RS - Portal do
  Estado do Rio Grande do Sul}.
\newblock
  \url{https://www.estado.rs.gov.br/com-investimento-inicial-de-r-3-bilhoes-governo-do-rs-e-scala-data-centers-assinam-acordo-para-o-maior-projeto-de-infraestrutu},
  Sep 2024.

\bibitem{scala2024pressrelease}
{Scala Datacenters}.
\newblock {Press Release - Enchentes Rio Grande do Sul}.
\newblock
  \footnotesize\url{https://www.scaladatacenters.com/doc/Press_Release_Enchentes_Rio_Grande_do_Sul_Scala_Data_Centers.pdf},
  May 2024.

\bibitem{sebrae-offline2024}
Sebrae.
\newblock {Devido a alagamento no Centro de POA, Sebrae RS desliga energia e
  Datacenter da Sede Metropolitana}.
\newblock
  \footnotesize\url{https://rs.agenciasebrae.com.br/dados/devido-a-alagamento-no-centro-de-poa-sebrae-rs-desliga-energia-e-datacenter-da-sede-metropolitana/},
  May 2024.

\bibitem{ixbr2024chuva}
Victor~Hugo Silva.
\newblock Chuvas no rs: uso de internet em porto alegre cai pela metade com
  obstáculos para acesso.
\newblock
  \footnotesize\url{https://g1.globo.com/tecnologia/noticia/2024/05/08/chuvas-no-rs-uso-de-internet-cai-pela-metade-com-obstaculos-para-acesso.ghtml},
  May 2024.

\bibitem{simet_escolas}
{SIMET}.
\newblock Escolas.
\newblock \footnotesize\url{https://conectividadenaeducacao.nic.br}, May 2024.

\bibitem{simet_saude}
{SIMET}.
\newblock Saúde.
\newblock \footnotesize\url{https://conectividadenasaude.nic.br}, May 2024.

\bibitem{shared2022simao}
Maileen~Schwarz Simão, Euler Ribeiro, Beatriz~Batista Cardoso, Marcos Aurélio
  Izumida~Martins, Moacir~Fernandes Lopes, and Fabricio Rodrigues.
\newblock Shared trench for burying cables in the conversion of overhead to
  underground networks.
\newblock In {\em 2022 IEEE Conference on Technologies for Sustainability},
  pages 120--124, 2022.

\bibitem{cloud_procergs}
Pamela Souza.
\newblock Exclusivo: Procergs migra dados para a nuvem em caráter de urgência
  por risco de inundação no rs - it forum.
\newblock
  \footnotesize\url{https://itforum.com.br/noticias/procergs-migra-dados-nuvem-risco-inundacao/},
  May 2024.

\bibitem{ripe2015atlas}
{Staff, RIPE NCC}.
\newblock {RIPE Atlas: A Global Internet Measurement Network}.
\newblock {\em Internet Protocol Journal}, 18(3), 2015.

\bibitem{brdigital2024teletime}
Teletime.
\newblock {No centro de Porto Alegre, a operação da BRDigital para manter
  dois datacenters funcionando}.
\newblock
  \footnotesize\url{https://teletime.com.br/08/05/2024/no-centro-de-porto-alegre-a-operacao-da-br-digital-para-manter-dois-datacenters-funcionando/},
  May 2024.

\bibitem{ObamaClimateDataInitiative}
{The Obama White House}.
\newblock {The President’s Climate Data Initiative: Empowering America’s
  Communities to Prepare for Climate Change}.
\newblock
  \url{https://obamawhitehouse.archives.gov/the-press-office/2014/03/19/fact-sheet-president-s-climate-data-initiative-empowering-america-s-comm},
  Mar 2014.

\bibitem{TenYearsNewOrleans}
{The Obama White House}.
\newblock {Ten Years After Katrina: New Orleans’ Recovery, and What Data Had
  to Do with it}.
\newblock
  \url{https://medium.com/@ObamaWhiteHouse/ten-years-after-katrina-new-orleans-recovery-and-what-data-had-to-do-with-it-3df0bb2467e9},
  Aug 2015.

\bibitem{cloud_tjrs}
{TJRS}.
\newblock Poder judiciário tribunal de justiça do estado do rio grande do
  sul.
\newblock
  \footnotesize\url{https://www.tjrs.jus.br/novo/noticia/migracao-para-a-nuvem-proporcionara-nova-experiencia-a-usuarios/},
  May 2024.

\bibitem{tornatore2016survey}
Massimo Tornatore, Joao Andr{\'e}, P{\'e}ter Babarczi, Torsten Braun, Eirik
  F{\o}lstad, Poul Heegaard, Ali Hmaity, Marija Furdek, Luisa Jorge, Wojciech
  Kmiecik, et~al.
\newblock A survey on network resiliency methodologies against weather-based
  disruptions.
\newblock In {\em 8th International Workshop on Resilient Networks Design and
  Modeling}, pages 23--34. IEEE, 2016.

\end{thebibliography}

\end{document}